\documentclass[aps,prl,nobibnotes,twocolumn,superscriptaddress,bibliography]{revtex4-2}
\usepackage{amsfonts}
\usepackage{mathrsfs}
\usepackage{amsmath}
\usepackage{color}
\usepackage{natbib}
\usepackage{graphicx}
\usepackage{bm}
\usepackage{amssymb}
\usepackage{xspace}
\usepackage{epstopdf}
\usepackage{dcolumn}
\usepackage{multirow}
\usepackage[colorlinks=true, letterpaper=true, pdfstartview=FitV, linkcolor=blue, citecolor=blue, urlcolor=blue]{hyperref}
\usepackage{wrapfig}
\usepackage[titletoc]{appendix}

\makeatletter

\newcommand{\Rmnum}[1]{\expandafter\@slowromancap\romannumeral #1@}
\makeatother

\begin{document}
\title{Quasi-One-Dimensional Spin Transport  in    Altermagnetic $Z^3$ Nodal Net Metals}

\author{Tingli He}\thanks{These authors contributed equally to this work.}
\affiliation{State Key Laboratory of Reliability and Intelligence of Electrical Equipment, School of Materials Science and Engineering, Hebei University of Technology, Tianjin 300401, China}
\affiliation{Key Lab of advanced optoelectronic quantum architecture and measurement (MOE), Beijing Key Lab of Nanophotonics $\&$ Ultrafine Optoelectronic Systems, and School of Physics, Beijing Institute of Technology, Beijing 100081, China}

\author{Lei Li}\thanks{These authors contributed equally to this work.}
\affiliation{Key Lab of advanced optoelectronic quantum architecture and measurement (MOE), Beijing Key Lab of Nanophotonics $\&$ Ultrafine Optoelectronic Systems, and School of Physics, Beijing Institute of Technology, Beijing 100081, China}

\author{Chaoxi Cui}
\affiliation{Key Lab of advanced optoelectronic quantum architecture and measurement (MOE), Beijing Key Lab of Nanophotonics $\&$ Ultrafine Optoelectronic Systems, and School of Physics, Beijing Institute of Technology, Beijing 100081, China}

\author{Run-Wu Zhang}
\affiliation{Key Lab of advanced optoelectronic quantum architecture and measurement (MOE), Beijing Key Lab of Nanophotonics $\&$ Ultrafine Optoelectronic Systems, and School of Physics, Beijing Institute of Technology, Beijing 100081, China}

\author{Zhi-Ming Yu}\email{zhiming\_yu@bit.edu.cn}
\affiliation{Key Lab of advanced optoelectronic quantum architecture and measurement (MOE), Beijing Key Lab of Nanophotonics $\&$ Ultrafine Optoelectronic Systems, and School of Physics, Beijing Institute of Technology, Beijing 100081, China}

\author{Guodong Liu}
\affiliation{State Key Laboratory of Reliability and Intelligence of Electrical Equipment, School of Materials Science and Engineering, Hebei University of Technology, Tianjin 300401, China}

\author{Xiaoming Zhang}\email{zhangxiaoming87@hebut.edu.cn}
\affiliation{State Key Laboratory of Reliability and Intelligence of Electrical Equipment, School of Materials Science and Engineering, Hebei University of Technology, Tianjin 300401, China}

\begin{abstract}
In three dimensions, quasi-one-dimensional (Q1D) transport has traditionally been associated with systems featuring a Q1D chain structure. Here, based on first-principle calculations, we  go beyond this  understanding to show that the Q1D transport can also be realized in certain three-dimensional (3D) altermagnetic (AM) metals with a topological  nodal net in momentum space  but lacking Q1D chain structure in real space, including the existing compounds $\beta$-Fe$_2$(PO$_4$)O, Co$_2$(PO$_4$)O, and LiTi$_2$O$_4$. These materials exhibit an AM ground state and feature an ideal crossed $Z^3$ Weyl nodal line in each spin channel around Fermi level, formed by three straight and flat nodal lines traversing the entire Brillouin zone. These nodal lines eventually lead to an AM $Z^3$ nodal net. Surprisingly, the electronic conductivity $\sigma_{xx}$ in these topological nodal net metals is dozens of times larger than $\sigma_{yy}$ and $\sigma_{zz}$ in the up-spin channel, while $\sigma_{yy}$ dominates transport in the down-spin channel. This suggests a distinctive Q1D transport signature in each spin channel, and the  principal moving directions for the two spin channels are orthogonal, resulting in Q1D direction-dependent spin transport.
This novel phenomenon cannot be found in both conventional 3D bulk materials and Q1D chain materials.
In particular, the Q1D spin transport gradually disappears as the Fermi energy moves away from the nodal net, further confirming its topological origin.
Our work not only enhances the comprehension of topological physics in altermagnets but also opens a new direction for the exploration of topological spintronics.
\end{abstract}
\maketitle

The coupling of magnetism and topological states has sparked extensive research interest \cite{PhysRevX.7.041069, tokura-magnetic-2019, xu-high-throughput-2020, elcoro-magnetic-2021, bernevig-progress-2022}. Magnetic topological materials exhibit a plethora of novel physical phenomena, including the quantum anomalous Hall effect \cite{PhysRevLett.123.096401,doi:10.1126/science.aax8156,PhysRevLett.127.236402}, anomalous Hall effect \cite{PhysRevLett.112.017205,PhysRevB.95.075128,li-giant-2020,PhysRevLett.129.097201}, anomalous Nernst effect \cite{guin2019zero,guin2019anomalous,PhysRevB.103.L201101,pan-giant-2022,PhysRevLett.129.097201}, and magnetoresistance effect \cite{suzuki2019singular,zhu-large-2023}, which hold promising application prospects in spintronics. Magnetic ordering, on one hand, diminishes the system's symmetry by breaking time-reversal symmetry (${\cal T}$) and certain crystal symmetries, presenting challenges in extending topological phases to magnetic systems \cite{yu2022encyclopedia,PhysRevB.105.104426,PhysRevB.105.085117,PhysRevB.103.115112}. On the other hand, it introduces an additional spin degree of freedom, paving the way for achieving electric control of spin, a critical pursuit in spintronics \cite{zhang2023predictable,  ma2021multifunctional,doi:10.1126/sciadv.adn0479, shaoantiferromagnetic2024, aelm.201800466, WOS:000527521300001, yan-electricfieldcontrolled-2020,  hu2022efficient, kim2022ferrimagnetic, chen2024emerging,  gong2023hidden}.

In recent years, significant progress has been made in realizing various topological phases within ferromagnetic  and antiferromagnetic  systems. Notable examples include PdF$_3$ \cite{PhysRevLett.129.097201}, Co$_3$Sn$_2$S$_2$ \cite{wanglarge2018, guin2019zero},  Co$_2$MnGa \cite{guin2019anomalous,PhysRevLett.119.156401}, and LiV$_2$F$_6$ \cite{doi:10.1021/acs.nanolett.1c02968},  Mn$_3$Sn/Ge \cite{PhysRevLett.119.087202}, CaCrO$_3$ \cite{PhysRevB.107.155126}, YbMnBi$_2$ \cite{pan-giant-2022},  etc. These magnetic compounds exhibit interesting  topological  states, such as Weyl points and nodal lines, along with anomalous  transport phenomena like the giant anomalous Hall effect and anomalous Nernst effect.

Remarkably, a new class of magnetic order: altermagnetism has been proposed very recently~\cite{doi:10.1126/sciadv.aaz8809, PhysRevLett.126.127701, PhysRevX.12.011028, PhysRevX.12.031042, PhysRevX.12.040501}, and has garnered considerable attention. The  altermagnetic (AM) systems feature  collinear-compensated magnetic order, and alternating spin polarizations in both crystal structure and electronic  band structure \cite{PhysRevX.12.031042}.
Particularly, due to large spin splitting and robustness against magnetic field perturbations, the AM materials  have demonstrated intriguing spin-dependent transport phenomena,  leading to AM spintronics \cite{PhysRevLett.126.127701,PhysRevX.12.011028, PhysRevLett.130.216702}.
Some topological phases have been achieved in   AM materials (they were recognized as  antiferromagnetic materials based on the previous classification of magnetic phases) \cite{PhysRevB.107.224402, PhysRevB.108.024410,  nag2024gdalsiantiferromagnetictopologicalweyl},
offering new material choices and operational mechanisms for the design and application of spintronic devices.

\begin{figure}
\includegraphics[width=8.8cm]{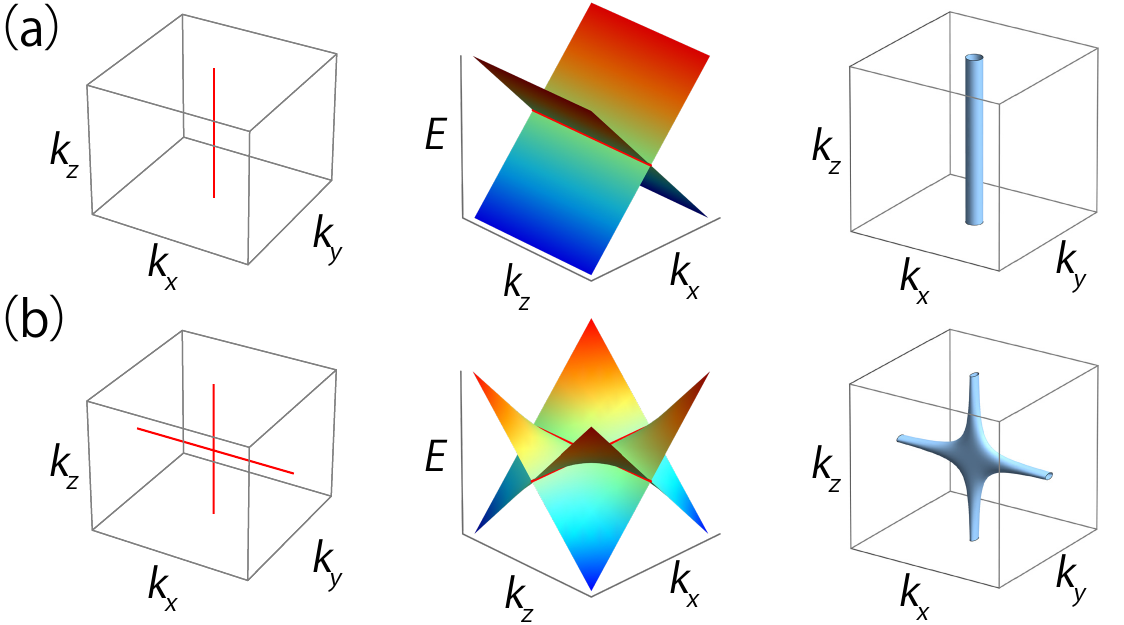}
\caption{Illustration of (a) an ideal $Z^3$ nodal line with $n_z=1$, and (b) an ideal crossed $Z^3$ nodal line with $n_z=1$ and $n_x=1$.
The figures from left to right denote the shape, the band structure and the Fermi surface of  the  nodal lines.
Since the Fermi surface of $Z^3$ and $Z^3$ crossed nodal lines respectively feature  cylindrical and flattened shapes, the two nodal lines should exhibit Q2D and Q1D transport properties.
\label{fig1}}
\end{figure}

A unique form of topological nodal line, termed the $Z^3$ nodal line, has been proposed in nonmagnetic systems \cite{PhysRevB.96.081106}. Unlike conventional nodal lines, this structure traverses the entire Brillouin zone (BZ) and is characterized by three integers $Z^3=(n_x,n_y,n_z)$, indicating the number of  times the line winds around each corresponding direction \cite{PhysRevB.96.081106,PhysRevB.99.121106}. Remarkably, owing to significant anisotropy, the ideal $Z^3$ nodal line is another structure besides layered and chain structures that can lead to low-dimensional transport phenomena. Here, the ``\emph{ideal}'' means the $Z^3$ nodal line is straight and has a small energy variation, as illustrated in Fig.~\ref{fig1}. For instance, an ideal $Z^3$ nodal line with $n_z=1$  would result in quasi-two-dimensional (Q2D) transport, with slower electron velocity along the $z$-direction compared to the $x$-$y$ plane [see Fig.~\ref{fig1}(a)].
With certain symmetries, two $Z^3$ nodal lines may form a crossed $Z^3$ nodal line, as shown in  Fig.~\ref{fig1}(b), where  the $y$-component longitudinal electronic conductivity will dominates the transport of system, suggesting a Q1D transport in topological crossed $Z^3$ nodal line semimetals.
However, unlike layered and chain structures defined in real space, the $Z^3$ and crossed $Z^3$ nodal line are  defined in momentum space.
Particularly, due to the altermagnetism, the crossed $Z^3$ nodal lines in AM materials generally come in pair with different $Z^3$ index, forming an AM $Z^3$ nodal net.
This is a unique property that cannot be found in ferromagnetic and  antiferromagnetic materials.
Therefore, a natural question then is: Does an ideal $Z^3$ nodal line or crossed $Z^3$ nodal line exist in AM materials?
If so, what novel transport phenomena distinguish them from the materials  with layered or chain structures.

In this work, we find that the AM materials: $\beta$-Fe$_2$(PO$_4$)O, Co$_2$(PO$_4$)O, and LiTi$_2$O$_4$, are ideal topological  $Z^3$  nodal net metals, featuring two crossed $Z^3$ nodal lines around Fermi level in both spin channels [see Fig.~\ref{fig3}(c)].
Previous studies have elucidated the electronic structures of  $\beta$-Fe$_2$(PO$_4$)O  and  LiTi$_2$O$_4$ \cite{he2019topological,PhysRevB.104.045143}. Additionally, $\beta$-Fe$_2$(PO$_4$)O has been identified as a X-type altermagnetic material \cite{zhang2023x}.
However, the impact of the AM $Z^3$ nodal net on the  transport properties of these systems remains unexplored.

Based on the first-principles calculations, we demonstrate that the  ideal AM  $Z^3$  nodal net indeed can generate  an unique  transport phenomenon, namely, in the up-spin and down-spin channels, the electronic conductivity around Fermi level are  dominated by $\sigma_{xx}^{\uparrow}$ and $\sigma_{yy}^{\downarrow}$, respectively, leading to novel Q1D direction-dependent spin transport. Notably, away from the AM $Z^3$  nodal net, the spin conductivity remains anisotropic but gradually loses its Q1D features, suggesting a close relationship between this novel spin transport and the AM $Z^3$ nodal net. 

\begin{figure}
\includegraphics[width=8.8cm]{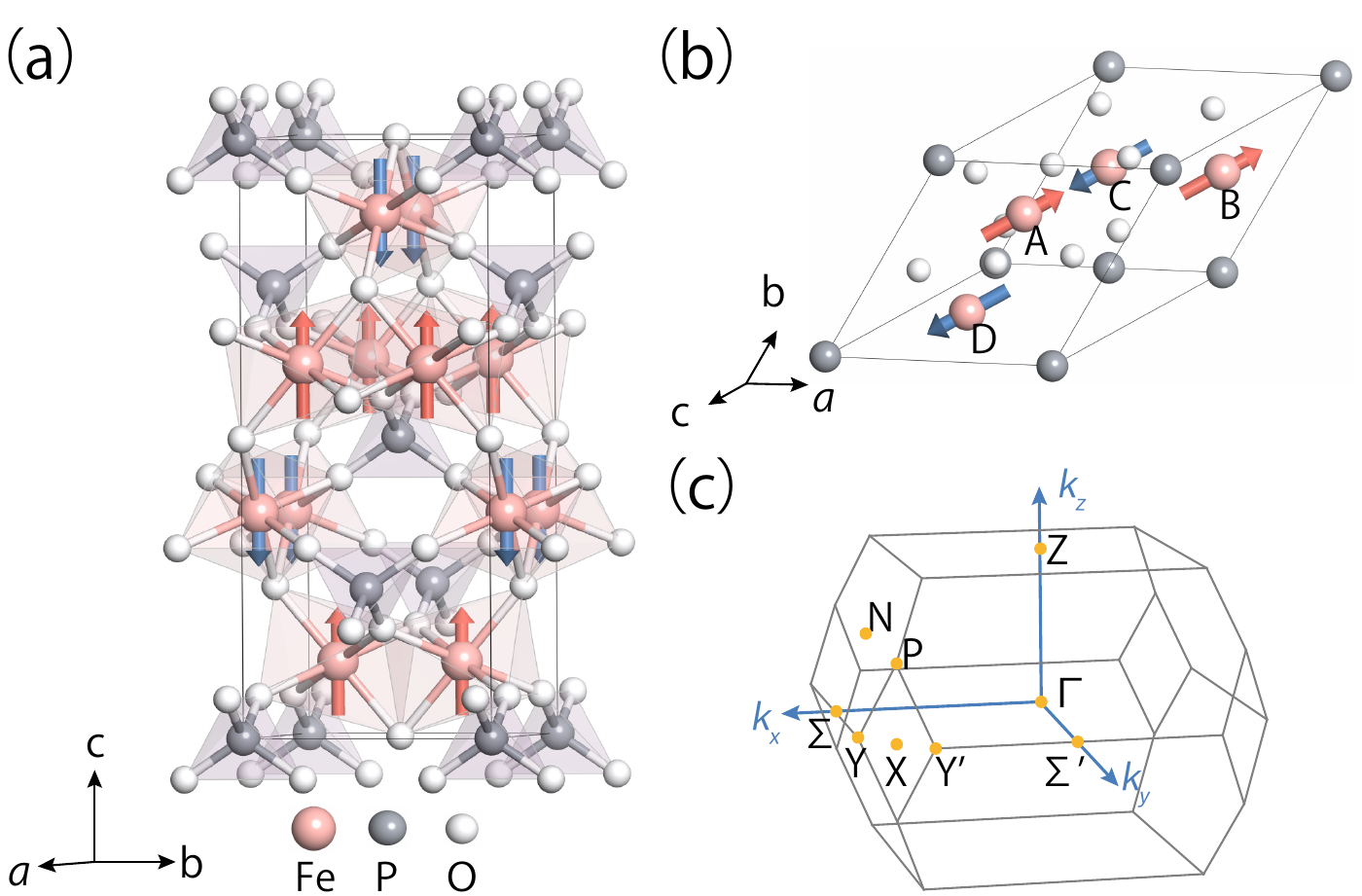}
\caption{ (a) The conventional  and (b) primitive cell of  $\beta$-Fe$_2$(PO$_4$)O. Red and blue arrows represent up-spin and down-spin magnetic moments, respectively. (c) denotes the  BZ.
\label{fig2}}
\end{figure}

\textit{\textcolor{blue}{Example 1: $\beta$-Fe$_2$(PO$_4$)O and Co$_2$(PO$_4$)O.}}--The material $\beta$-Fe$_2$(PO$_4$)O is a mixed-valence iron oxyphosphate featuring Fe$^{2+}$/Fe$^{3+}$ oxidation states \cite{MODARESSI1981301,ECHCHAHED198847,IJJAALI1990195}.
It has been experimentally identified as an antiferromagnetic material with a N\'eel temperature as high as 408 K \cite{IJJAALI1990195} in the 1980s, but according to the modern classification for magnetic systems  \cite{PhysRevX.12.031042}, it is an AM material \cite{zhang2023x}.

The crystal structure of $\beta$-Fe$_2$(PO$_4$)O is depicted Fig.~\ref{fig2}(a-b), showing it is not a typical layered or chain material.
Fe and P atoms occupy the 8$d$ and 4$a$ Wyckoff sites, respectively, while O atoms are distributed across the 4$b$ and 16$h$ Wyckoff sites.
Magnetic moments primarily reside on Fe sites with an approximate magnitude of 4 $\mu_B$, and the N\'eel vector is along the $z$ axis [see Fig.~\ref{fig2}(a)], consistent with previous calculations \cite{he2019topological}. The optimized lattice parameters of the conventional cell are found to be $a = b = 5.419$ Å and $c = 12.657$ Å, in good agreement with experimental values \cite{IJJAALI1990195}.
The magnetic space group (MSG) for $\beta$-Fe$_2$(PO$_4$)O is obtained as No. 141.554 ($I4'_1$/$am'd$) \cite{stokes2005findsym}.
This MSG lacks four-fold rotation $\tilde{{\cal C}}_{4z}$,  time-reversal symmetry  $\mathcal{T}$, and ${\cal P}{\mathcal T}$ (${\cal P}$ being inversion symmetry), but exhibits the combined operator $\tilde{{\cal C}}_{4z}\mathcal{T}$, which connects the  up-spin and down-spin channels  in $\beta$-Fe$_2$(PO$_4$)O.
Thus, the $\beta$-Fe$_2$(PO$_4$)O will exhibit spin splitting in a generic point in the BZ, reflecting the AM character \cite{PhysRevX.12.031042, PhysRevX.12.040501}.
Moreover,  while the   electronic conductivity of $\beta$-Fe$_2$(PO$_4$)O in each spin channel is generally  anisotropic, it always has $\sigma_{xx}^{\uparrow}=\sigma_{yy}^{\downarrow}$ and  $\sigma_{zz}^{\uparrow}=\sigma_{zz}^{\downarrow}$ due to   $\tilde{{\cal C}}_{4z}\mathcal{T}$.

\begin{figure}
	\includegraphics[width=8.8cm]{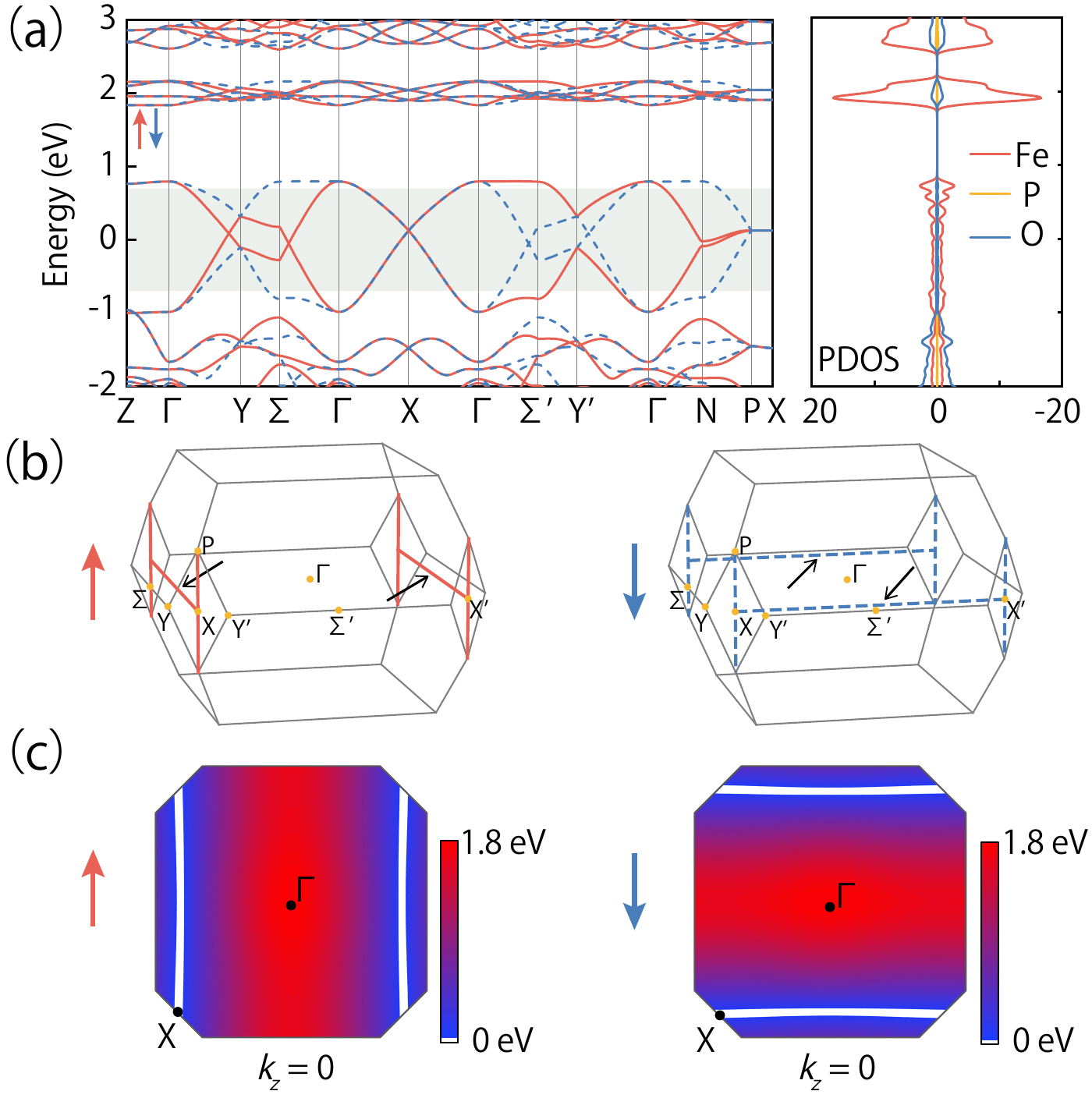}
	\caption{(a) Electronic bands  and  projected density of states (PDOS) of $\beta$-Fe$_2$(PO$_4$)O.
		The energy region where the $Z^3$ AM nodal net appears is highlighted by  green.
	(b) The  shapes of the $Z^3$ crossed nodal lines in up-spin and down-spin channels. The two $Z^3$ crossed nodal lines form a  $Z^3$ nodal net. (c) The shape of the $Z^3$ nodal line on the $k_z$ = 0 plane for each spin channel indicated by the  arrows in (b). 	The color map indicates the local gap between the two bands that construct the  $Z^3$ nodal line.
		\label{fig3}}
\end{figure}

Since the  spin-orbit coupling (SOC) in  $\beta$-Fe$_2$(PO$_4$)O  is negligible \cite{SM}, we consider the spin-resolved band structure of the material without SOC, which is plotted  in Fig.~\ref{fig3}(a).
Interestingly, in each spin channel, there exist  two bands around the Fermi level that are  degenerate along two inequivalent $P$-$X$ paths,  resulting in two ideal $Z^3$ nodal lines with $n_z=1$, as the lines  are not only straight but also flat [see Fig.~\ref{fig3}(a,b)].
Moreover, in the $k_z=0$ plane, $\beta$-Fe$_2$(PO$_4$)O features a $Z^3$ Weyl nodal line in both  spin channels, protected by the glide mirror symmetries $\tilde{{\cal M}}_{z}$. The profiles of the Weyl nodal lines are numerically calculated and  depicted in Fig.~\ref{fig3}(c), showing that the Weyl nodal lines in  both  spin channels are very straight and traverse the BZ.
Thus, the up-spin (down-spin) channel exhibits a $Z^3$ crossed nodal line with $n_z=1$ and $n_y=1$ ( $n_x=1$), as depicted in Fig.~\ref{fig3}(b).
Furthermore, the  two spin-resolved $Z^3$ crossed nodal lines contact with each other at the $X$ point in the BZ,  ultimately forming an AM $Z^3$ nodal net. A detailed symmetry analysis of such a $Z^3$ nodal net is provided in the  Supplemental Material (SM) \cite{SM}.

\begin{figure}
	\includegraphics[width=8.8cm]{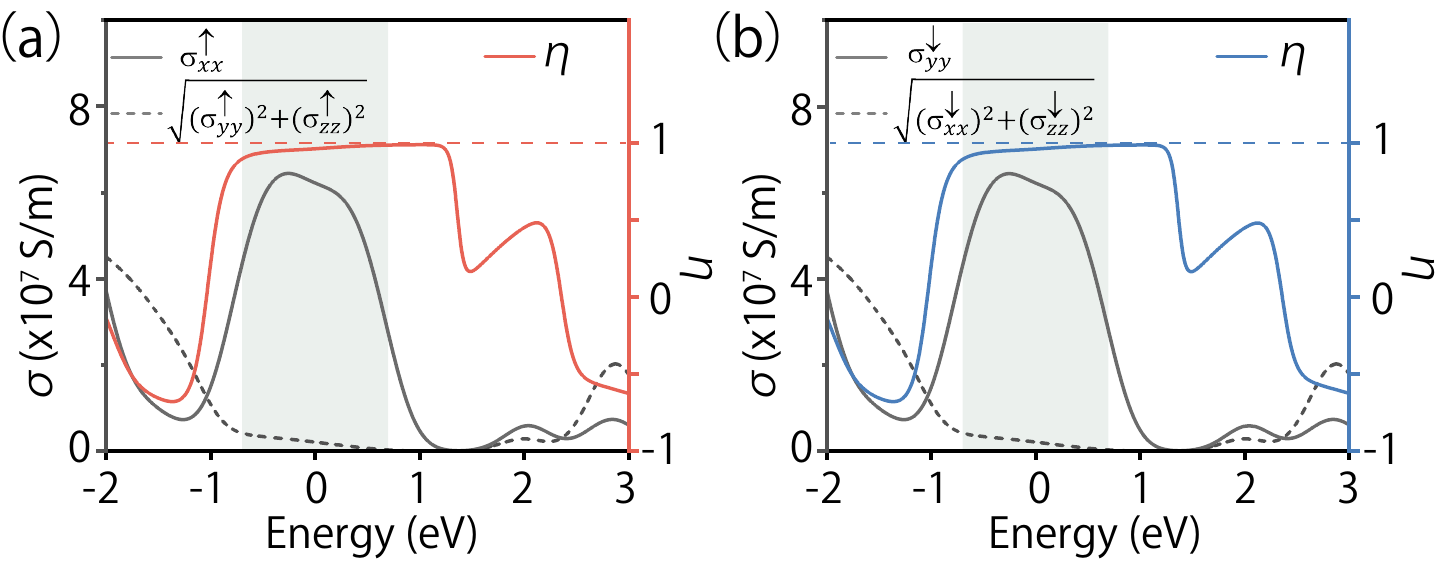}
	\caption{The calculated electronic conductivity (left axis) and  $\eta$ (right axis) of  the $\beta$-Fe$_2$(PO$_4$)O compound in (a) up-spin  and (b) down-spin channels. The $Z^3$ AM nodal net appears in the green shaded region. Here, we set temperature of system as $T=50$ K. Hence, the  electronic conductivity is finite in the band gap.
		\label{fig4}}
\end{figure}

As all the $Z^3$ Weyl nodal lines in $\beta$-Fe$_2$(PO$_4$)O are ideal, one can expect that the longitudinal conductivity $\sigma_{xx}^{\uparrow}$ ($\sigma_{yy}^{\downarrow}$) of the up-spin (down-spin) channel in this material should be much larger than $\sigma_{yy}^{\uparrow}$ ($\sigma_{xx}^{\downarrow}$) and $\sigma_{zz}^{\uparrow}$ ($\sigma_{zz}^{\downarrow}$).
In the absence of SOC, the electronic conductivity of up-spin and down-spin electrons can be accurately and independently calculated. Therefore, to investigate the unique Q1D direction-dependent spin transport properties of the AM $Z^3$ nodal net in $\beta$-Fe$_2$(PO$_4$)O, we calculate the spin-resolved longitudinal electronic conductivities for both spin channels. The obtained results are shown in  Fig.~\ref{fig4}, from which three key features are observed.

Firstly,  the electronic conductivity in each spin generally is  anisotropic that $\sigma_{xx}^{\uparrow(\downarrow)}\neq\sigma_{yy}^{\uparrow(\downarrow)}\neq \sigma_{zz}^{\uparrow(\downarrow)}$.
However,  the electronic conductivities for the two spin channels are not independent, but have $\sigma_{xx}^\uparrow=\sigma_{yy}^\downarrow$, $\sigma_{yy}^\uparrow=\sigma_{xx}^\downarrow$, and  $\sigma_{zz}^\uparrow=\sigma_{zz}^\downarrow$, consistent with the symmetry analysis.

Secondly, we find that the anisotropy of the electronic  conductivity in each spin channel is  extremely enhanced around the AM nodal net (Fermi level),  where $\sigma_{xx}^{\uparrow}$ ($\sigma_{yy}^{\downarrow}$) is dozens of times larger than the other two longitudinal conductivities.
To quantitatively describe it, we define a  dimensionless quantity
\begin{equation}
	\eta  = \frac{\sigma_{xx}^{\uparrow}-\sigma_{\perp}^{\uparrow}}{\sqrt{(\sigma_{xx}^\uparrow)^2+(\sigma_{\perp}^{\uparrow})^2}} =\frac{\sigma_{yy}^{\downarrow}-\sigma_{\perp}^{\downarrow}}{\sqrt{(\sigma_{yy}^\downarrow)^2+(\sigma_{\perp}^{\downarrow})^2}},
\end{equation}
with $\sigma_\perp^{\uparrow(\downarrow)}={\sqrt{(\sigma_{yy(xx)}^{\uparrow(\downarrow)})^2+(\sigma_{zz}^{\uparrow(\downarrow)})^2}}$.
$\eta=1$ corresponds to the strongest anisotropy and  ideal  Q1D direction-dependent  spin transport, where $\sigma_{xx}^{\uparrow}$ and $\sigma_{yy}^{\downarrow}$ completely  dominate the electronic transport in  up-spin and down-spin channels, respectively.
As shown in  Fig. \ref{fig4}, $\eta$ approaches 1 only in the energy range where the AM nodal net appears.
We have further checked that in this energy range, the Fermi surface of the system indeed has a flattened  shape \cite{SM}.

Remarkably, the  Q1D  transport in $\beta$-Fe$_2$(PO$_4$)O is distinct from systems with chain structures. For the former, although the spin current has Q1D characteristic, the electric current is isotropic in the $x$-$y$ plane, as $\sigma_{xx}^\uparrow+\sigma_{xx}^\downarrow=\sigma_{yy}^\uparrow+\sigma_{yy}^\downarrow$. In contrast, for the latter, both  electric and spin currents exhibit Q1D signature.
Moreover, due to the direction-dependent  Q1D  spin transport, $\beta$-Fe$_2$(PO$_4$)O can be utilized to generate a current with strong spin polarization, and the spin polarization can be easily switched by rotating the sample 90$^\circ$ along  $z$ axis.

Thirdly,  the PDOS in  Fig.~\ref{fig3}(a) shows that the  energy bands from -1 eV to 3 eV are mainly contributed by the Fe atoms.
However, away from the $Z^3$ AM nodal net, the spin conductivity  loses the Q1D signature, as $\eta$ rapidly deviates from 1 in the lower energy ($<-0.8$ eV), and also is far from 1 in the higher energy ($> 1.8$ eV), as shown in  Fig.~\ref{fig4}. This strongly suggests that the spin-polarized Q1D transport is closely associated with the $Z^3$ nodal net rather than the  structural arrangement  of the atoms in  $\beta$-Fe$_2$(PO$_4$)O.

The synthesis of Co$_2$(PO$_4$)O has also been reported \cite{doi:10.1021/ic4029904}, showing a similar crystalline structure, magnetic configuration,  electronic bands and the Q1D spin transport as $\beta$-Fe$_2$(PO$_4$)O \cite{SM}.

\begin{figure}
\includegraphics[width=8.8cm]{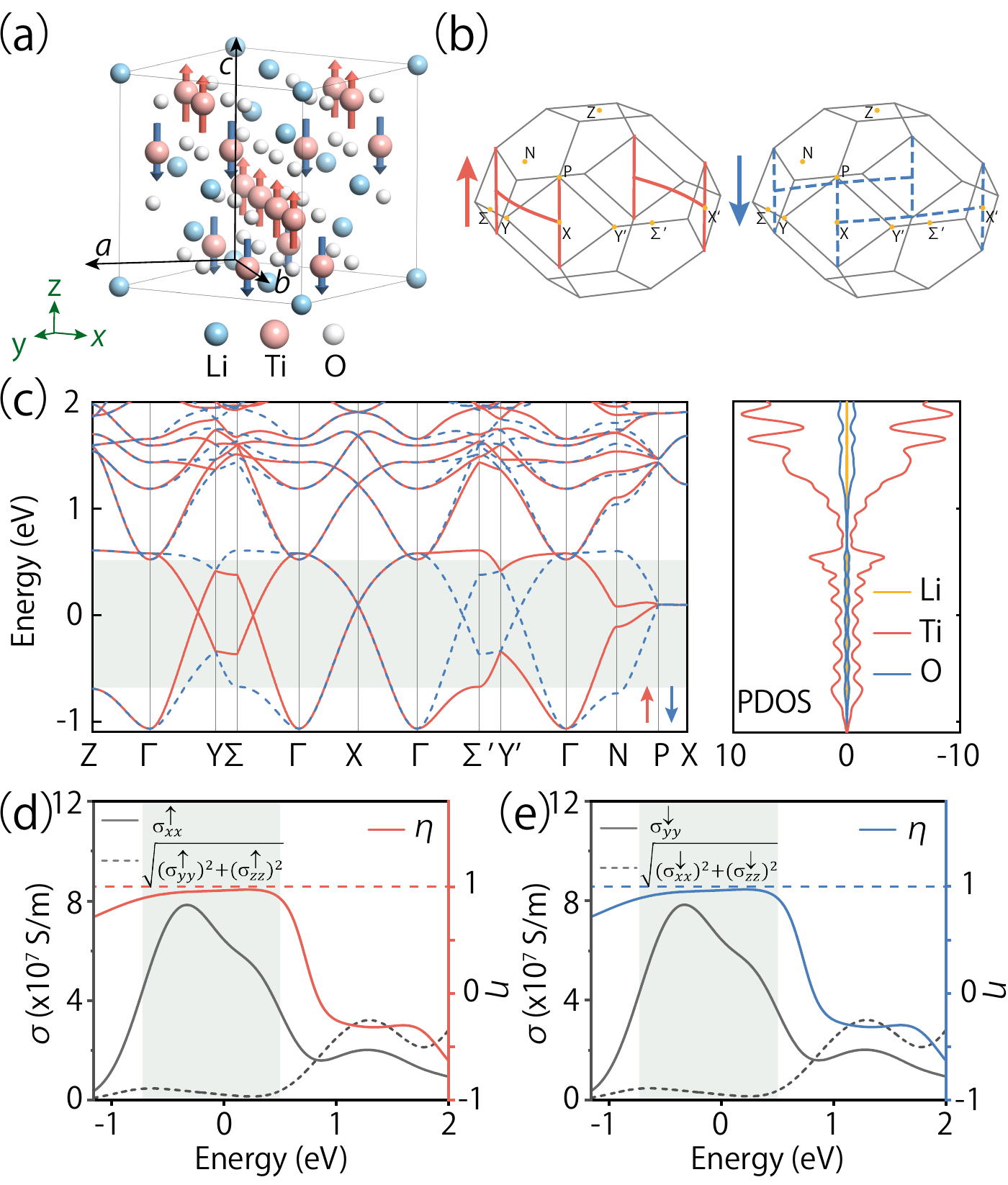}
\caption{(a) Crystal structures of  LiTi$_2$O$_4$. Red and blue arrows represent up-spin and down-spin magnetic moments, respectively. (b) BZ and the shape of the $Z^3$ crossed  nodal line in each spin channel.  (c) Electronic bands and  PDOS of LiTi$_2$O$_4$. (d,e) The calculated electronic conductivity (left axis)  and  $\eta$ (right axis) in (d) up-spin  and (e) down-spin channels.	The $Z^3$ AM nodal net appears in the green shaded region.
  \label{fig5}}
\end{figure}

\textit{\textcolor{blue}{Example 2: LiTi$_2$O$_4$.}}--The LiTi$_2$O$_4$ also exhibits an AM $Z^3$ nodal net near the Fermi level.
The lattice structure of LiTi$_2$O$_4$ without magnetic momentum belongs to a cubic Bravais lattice, which differs from that of $\beta$-Fe$_2$(PO$_4$)O.
However, the ground state of LiTi$_2$O$_4$ has AM configuration, as  depicted in Fig.~\ref{fig5}(a).
Since the  N\'eel vector is incompatible with  the crystal symmetry, the MSG of the LiTi$_2$O$_4$ is calculated as  No. 141.554 \cite{stokes2005findsym,PhysRevB.104.045143}.
The  optimized lattice parameters of the conventional cell are $a = b = 6.053 $ Å and $c = 8.753 $ Å.

Figure~\ref{fig5}(c) shows the electronic bands of the LiTi$_2$O$_4$ without SOC, as SOC effect here is  negligible. Scanning through the whole BZ reveals the existence  of an ideal AM $Z^3$ nodal net [see Fig.~\ref{fig5}(b)], which share similar configuration as that in  $\beta$-Fe$_2$(PO$_4$)O.
Particularly, the  direction-dependent Q1D spin transport is observed in LiTi$_2$O$_4$, and again only appears in the energy where AM $Z^3$ nodal net appears [see
 Fig.~\ref{fig5}(c-e)]. This further confirms that Q1D spin transport is a common characteristic of materials possessing an ideal $Z^3$ nodal net.

\textit{\textcolor{blue}{Lattice model of AM $Z^3$ nodal net.}}--
To further understand the Q1D spin transport, we establish a simple effective lattice model for the AM $Z^3$ nodal net, based on $\beta$-Fe$_2$(PO$_4$)O.
We consider a 3D tetragonal body-centred lattice with an AM ordering, and assume that this lattice has the same MSG (No. 141.554) as the $\beta$-Fe$_2$(PO$_4$)O.
The unit cell of the system  includes an even number of lattice sites: up spins on half of sites and down spins on the remaining half.
The lattice has four atoms with $d_{x^2-y^2}$-like orbitals \cite{SM} per unit cell at  $8d$ Wyckoff sites (labeled as $\{A,B,C,D\}$), which correspond to the positions occupied  by the Fe  atoms of $\beta$-Fe$_2$(PO$_4$)O {[}see Fig. \ref{fig2}(b){]}.

The lattice belongs to SG No. 141, which is generated by a four-fold screw-rotation $\tilde{{\cal C}}_{4z}$ = $\{C_{4z}|\frac{1}{4}\frac{-1}{4}\frac{1}{4}\}$, a two-fold rotation ${C}_{2x}$, and  spatial inversion ${\cal P}$. Without altermagnetic ordering, the lattice also has time-reversal symmetry ${\cal T}$.
In the absence of SOC, using the four sites as a basis, the symmetry operators can be represented as
$\tilde{{\cal C}}_{4z} = (\Gamma_{+,0}+ \Gamma_{-,1})/2$, $C_{2x} =( \Gamma_{0,0}+\Gamma_{3,0}+\Gamma_{0,1} - \Gamma_{3,1})/2$,   ${\cal{P}}=\Gamma_{0,0}$, and ${\cal T}=\Gamma_{0,0}{\cal K}$ with  $\Gamma_{i,j}=\sigma_{i}\otimes\sigma_{j}$ ($i,j=0,1,2,3,\pm$), Here, $\boldsymbol{\sigma}$ denote the Pauli matrixes, $\sigma_{0}$ is the  identity matrix, and $\sigma_{\pm} = \sigma_{1}  \pm i\sigma_{2}$.
The corresponding model Hamiltonian under these symmetry constraints may be written as \cite{ZHANG2022108153,zhang2023magnetickp}
\begin{eqnarray}
	{\cal H}_{0} & =\left[\begin{array}{cc}
		H_1 & H' \\
		H'^{\dagger} & H_2
	\end{array}\right],
\end{eqnarray}
where $H_1=t_1\cos\frac{k_x+k_z}{2} \sigma_{1}$, $H_2=t_1\cos\frac{k_y+k_z}{2} \sigma_{1}$ and
\begin{eqnarray}
	H' & =t_2\left[\begin{array}{cc}
		\cos\frac{k_z}{2} & \cos\frac{k_y}{2} \\
		\cos\frac{k_x}{2} & \cos\frac{k_x+k_y+k_z}{2}
	\end{array}\right],
\end{eqnarray}
with $t_{i}$ ($i=1,2$) the real parameters.

Then, we introduce the spin degree of freedom and turn on a AM ordering along $z$-direction {[}see Fig. \ref{fig2}(a){]}. The AM ordering breaks ${\cal T}$, $\tilde{{\cal C}}_{4z}$  and $C_{2x}$, while holds $\tilde{{\cal C}}_{2z}$, ${\cal{P}}$  and the combined operation $\tilde{{\cal M}}_{z}={\cal{P}}\tilde{{\cal C}}_{2z}$, $\tilde{{\cal C}}_{4z}{\cal T}$ and $C_{2x}{\cal T}$.
To the leading order, the AM Hamiltonian may be written as \cite{PhysRevB.96.081107}
\begin{eqnarray}
	{\cal H} & = & {\cal H}_{0}s_{0}+J_{z}s_{z}\Gamma_{3,0},\label{eq:H}
\end{eqnarray}
with $\boldsymbol{s}$ the Pauli matrixes acting on spin space and $J_{z}$ denoting the strength of the AM potential.
When the AM potential ($J_z>0$) is much stronger than other hopping and energy terms, the eight bands are divided into two groups with each group including four bands.
Since the two groups are separated in energy space by $2J_{z}$, the model (\ref{eq:H}) can be simplified as two effectively four-band model \cite{PhysRevB.95.115138}.
The basis of the upper subsystem is $(|A\uparrow\rangle,|B\uparrow\rangle,|C\downarrow\rangle,|D\downarrow\rangle)^{T}$ and that for the lower subsystem is $(|A\downarrow\rangle,|B\downarrow\rangle,|C\uparrow\rangle,|D\uparrow\rangle)^{T}$.
We focus on the upper subsystem  and omit $J_{z}$  from it, as it only shifts the bands.
Then  the upper four-band model can be expressed as \cite{winkler2003spin}
\begin{align}
	{\cal H}_{\rm {eff}} &=\frac{t^2_2}{4J_z}+\left[\begin{array}{cc}
		H_1 & 0 \\
		0 & H_2
	\end{array}\right]+\frac{t^2_2}{4J_z}
	\left[\begin{array}{cc}
		H_1^{\prime} & 0 \\
		0 & H_2^{\prime}
	\end{array}\right], \label{eq:Heff}
\end{align}
with
\begin{align}
	H_1^{\prime}({\bm k}) &=2(\cos\frac{k_x}{2}\cos\frac{k_z}{2}+\cos\frac{k_y}{2}\cos\frac{k_x+k_y+k_z}{2})\sigma_1 \nonumber  \\
	&+\left[\begin{array}{cc}
		\cos k_y+\cos k_z & 0 \\
		0 & \cos k_x+\cos(k_x+k_y+k_z)
	\end{array}\right],\nonumber
\end{align}
and $H_2^{\prime}(k_x,k_y,k_z)=H_1^{\prime}(k_y,k_x,k_z)$.

The  band structure of ${\cal H}_{\rm {eff}} $ is calculated and plotted in Ref. \cite{SM}, which qualitatively reproduces the band structure and the AM $Z^3$ nodal net of the $\beta$-Fe$_2$(PO$_4$)O, as well as the Q1D spin transport.
Moreover, we  find that the Q1D spin transport will disappear when the AM $Z^3$ nodal net is removed \cite{SM}.
Our simple four-band effective model not only further demonstrate the closed relationship between AM $Z^3$ nodal net and Q1D spin transport, but also can serve as a good starting point for the subsequent  study of the physics of AM $Z^3$ nodal net.

\textit{\textcolor{blue}{Conclusions.}}--We have demonstrated the feasibility of achieving  ideal AM $Z^3$ nodal net in real materials and shown that the novel Q1D spin transport phenomena inherent to this topological structure. We have detailed the mechanisms whereby electrons with up-spin predominantly flow in one direction, while those with down-spin move primarily in a perpendicular direction, a characteristic distinctly contrasting with conventional 3D bulk materials and Q1D chain materials. This behavior could be effectively detected through the measurement of anisotropic spin resistance, offering a practical approach to exploring the unique physics of AM systems. Our work not only provides a solid foundation for exploring the novel physics of AM systems but also holds great promise for applications in topological spintronics.

\acknowledgments
This work is supported by the National Natural Science Foundation of China (Grants Nos. 12004035, 12274112, 12347117 and 12304188) and the China Postdoctoral Science Foundation (Grants Nos. 2024M754079 and GZC20242181). X. Zhang also thanks the sponsorship from S$\&T$ Program of Hebei (225676163GH).

\textit{\textcolor{blue}{Appendix on numerical methods.}}--
In this work, the electronic conductivity $\sigma(\mu, T)$ was calculated using the Boltzmann transport equation \cite{PIZZI2014422}. The conductivities for the two spin channels have a strong dependence on the chemical potential  $\mu$ and temperature $T$, which can be calculated as follows  \cite{ziman1972principles, PhysRevB.68.125210}:
\begin{equation}
	\sigma_{ij}(\mu, T) = e^2 \int_{-\infty}^{+\infty} dE \left[ - \frac{\partial f(E, \mu, T)}{\partial E} \right] \Sigma_{ij}(E),
\end{equation}
where $f(E, \mu, T)$ is the usual Fermi-Dirac distribution function, defined as
\begin{equation}
	f(E, \mu, T) = \frac{1}{e^{(E - \mu)/k_B T} + 1},
\end{equation}
and $\Sigma_{ij}(\epsilon)$ is the transport distribution function tensor, expressed as
\begin{equation}
	\Sigma_{ij}(\epsilon) = \frac{1}{N_k \Omega_c} \sum_{n, \mathbf{k}} v_i(n, \mathbf{k}) v_j(n, \mathbf{k}) \tau(n, \mathbf{k}) \delta(E - E_{n,\mathbf{k}}),
\end{equation}
where $n$ is the band index, $E_{n,\mathbf{k}}$ is energy of the $n$-th band at momentum point  $\mathbf{k}$, $v_i(n, \mathbf{k})$ denotes the $i$-th component of the velocity of the $n$-th band at $\mathbf{k}$,  $\delta$ stands for Dirac's delta function, and $\tau(n, \mathbf{k})$ is the relaxation time.  $N_k$ and $\Omega_c$ represent the number of $k$-points used to sample the Brillouin zone and the unit cell volume, respectively. In this work, the  relaxation time approximation was adopted \cite{PhysRevB.68.125210,madsen2006automated,nag2012electron}, in such case $\tau$ is independent of $n$ and $\mathbf{k}$.

For  $\beta$-Fe$_2$(PO$_4$)O, the electronic conductivity was calculated at a temperature of 50 K, which is much lower than N\'eel temperature of the system  $T_N$ = 408 K, using a $100 \times 100 \times 100$  momentum mesh via the Wannier90 code \cite{mostofi2008wannier90, mostofi2014updated}.
 Since the  relaxation time of the typical AM metal RuO$_2$ is theoretically estimated as $\sim$ 600 fs  \cite{PhysRevX.12.011028, feng2022anomalous}, we set the relaxation time of $\beta$-Fe$_2$(PO$_4$)O as a constant of 500 fs, which is comparable with that of RuO$_2$.
Similarly, for LiTi$_2$O$_4$ and Co$_2$(PO$_4$)O compounds,  $T= 50 $ K,  $\tau = 500 $ fs, and  a momentum mesh  of  $100 \times 100 \times 100$ are  used to calculate the  electronic conductivity.

\bibliography{ref}

\begin{thebibliography}{82}%
\makeatletter
\providecommand \@ifxundefined [1]{%
 \@ifx{#1\undefined}
}%
\providecommand \@ifnum [1]{%
 \ifnum #1\expandafter \@firstoftwo
 \else \expandafter \@secondoftwo
 \fi
}%
\providecommand \@ifx [1]{%
 \ifx #1\expandafter \@firstoftwo
 \else \expandafter \@secondoftwo
 \fi
}%
\providecommand \natexlab [1]{#1}%
\providecommand \enquote  [1]{``#1''}%
\providecommand \bibnamefont  [1]{#1}%
\providecommand \bibfnamefont [1]{#1}%
\providecommand \citenamefont [1]{#1}%
\providecommand \href@noop [0]{\@secondoftwo}%
\providecommand \href [0]{\begingroup \@sanitize@url \@href}%
\providecommand \@href[1]{\@@startlink{#1}\@@href}%
\providecommand \@@href[1]{\endgroup#1\@@endlink}%
\providecommand \@sanitize@url [0]{\catcode `\\12\catcode `\$12\catcode
  `\&12\catcode `\#12\catcode `\^12\catcode `\_12\catcode `\%12\relax}%
\providecommand \@@startlink[1]{}%
\providecommand \@@endlink[0]{}%
\providecommand \url  [0]{\begingroup\@sanitize@url \@url }%
\providecommand \@url [1]{\endgroup\@href {#1}{\urlprefix }}%
\providecommand \urlprefix  [0]{URL }%
\providecommand \Eprint [0]{\href }%
\providecommand \doibase [0]{https://doi.org/}%
\providecommand \selectlanguage [0]{\@gobble}%
\providecommand \bibinfo  [0]{\@secondoftwo}%
\providecommand \bibfield  [0]{\@secondoftwo}%
\providecommand \translation [1]{[#1]}%
\providecommand \BibitemOpen [0]{}%
\providecommand \bibitemStop [0]{}%
\providecommand \bibitemNoStop [0]{.\EOS\space}%
\providecommand \EOS [0]{\spacefactor3000\relax}%
\providecommand \BibitemShut  [1]{\csname bibitem#1\endcsname}%
\let\auto@bib@innerbib\@empty
\bibitem [{\citenamefont {Kruthoff}\ \emph {et~al.}(2017)\citenamefont
  {Kruthoff}, \citenamefont {de~Boer}, \citenamefont {van Wezel}, \citenamefont
  {Kane},\ and\ \citenamefont {Slager}}]{PhysRevX.7.041069}%
  \BibitemOpen
  \bibfield  {author} {\bibinfo {author} {\bibfnamefont {J.}~\bibnamefont
  {Kruthoff}}, \bibinfo {author} {\bibfnamefont {J.}~\bibnamefont {de~Boer}},
  \bibinfo {author} {\bibfnamefont {J.}~\bibnamefont {van Wezel}}, \bibinfo
  {author} {\bibfnamefont {C.~L.}\ \bibnamefont {Kane}},\ and\ \bibinfo
  {author} {\bibfnamefont {R.-J.}\ \bibnamefont {Slager}},\ }\bibfield  {title}
  {\bibinfo {title} {{Topological Classification of Crystalline Insulators
  through Band Structure Combinatorics}},\ }\href
  {https://doi.org/10.1103/PhysRevX.7.041069} {\bibfield  {journal} {\bibinfo
  {journal} {Phys. Rev. X}\ }\textbf {\bibinfo {volume} {7}},\ \bibinfo {pages}
  {041069} (\bibinfo {year} {2017})}\BibitemShut {NoStop}%
\bibitem [{\citenamefont {Tokura}\ \emph {et~al.}(2019)\citenamefont {Tokura},
  \citenamefont {Yasuda},\ and\ \citenamefont
  {Tsukazaki}}]{tokura-magnetic-2019}%
  \BibitemOpen
  \bibfield  {author} {\bibinfo {author} {\bibfnamefont {Y.}~\bibnamefont
  {Tokura}}, \bibinfo {author} {\bibfnamefont {K.}~\bibnamefont {Yasuda}},\
  and\ \bibinfo {author} {\bibfnamefont {A.}~\bibnamefont {Tsukazaki}},\
  }\bibfield  {title} {\bibinfo {title} {{Magnetic topological insulators}},\
  }\href {https://doi.org/10.1038/s42254-018-0011-5} {\bibfield  {journal}
  {\bibinfo  {journal} {Nat. Rev. Phys.}\ }\textbf {\bibinfo {volume} {1}},\
  \bibinfo {pages} {126} (\bibinfo {year} {2019})}\BibitemShut {NoStop}%
\bibitem [{\citenamefont {Xu}\ \emph {et~al.}(2020)\citenamefont {Xu},
  \citenamefont {Elcoro}, \citenamefont {Song}, \citenamefont {Wieder},
  \citenamefont {Vergniory}, \citenamefont {Regnault}, \citenamefont {Chen},
  \citenamefont {Felser},\ and\ \citenamefont
  {Bernevig}}]{xu-high-throughput-2020}%
  \BibitemOpen
  \bibfield  {author} {\bibinfo {author} {\bibfnamefont {Y.}~\bibnamefont
  {Xu}}, \bibinfo {author} {\bibfnamefont {L.}~\bibnamefont {Elcoro}}, \bibinfo
  {author} {\bibfnamefont {Z.-D.}\ \bibnamefont {Song}}, \bibinfo {author}
  {\bibfnamefont {B.~J.}\ \bibnamefont {Wieder}}, \bibinfo {author}
  {\bibfnamefont {M.~G.}\ \bibnamefont {Vergniory}}, \bibinfo {author}
  {\bibfnamefont {N.}~\bibnamefont {Regnault}}, \bibinfo {author}
  {\bibfnamefont {Y.}~\bibnamefont {Chen}}, \bibinfo {author} {\bibfnamefont
  {C.}~\bibnamefont {Felser}},\ and\ \bibinfo {author} {\bibfnamefont {B.~A.}\
  \bibnamefont {Bernevig}},\ }\bibfield  {title} {\bibinfo {title}
  {{High-throughput calculations of magnetic topological materials}},\ }\href
  {https://doi.org/10.1038/s41586-020-2837-0} {\bibfield  {journal} {\bibinfo
  {journal} {Nature}\ }\textbf {\bibinfo {volume} {586}},\ \bibinfo {pages}
  {702} (\bibinfo {year} {2020})}\BibitemShut {NoStop}%
\bibitem [{\citenamefont {Elcoro}\ \emph {et~al.}(2021)\citenamefont {Elcoro},
  \citenamefont {Wieder}, \citenamefont {Song}, \citenamefont {Xu},
  \citenamefont {Bradlyn},\ and\ \citenamefont
  {Bernevig}}]{elcoro-magnetic-2021}%
  \BibitemOpen
  \bibfield  {author} {\bibinfo {author} {\bibfnamefont {L.}~\bibnamefont
  {Elcoro}}, \bibinfo {author} {\bibfnamefont {B.~J.}\ \bibnamefont {Wieder}},
  \bibinfo {author} {\bibfnamefont {Z.}~\bibnamefont {Song}}, \bibinfo {author}
  {\bibfnamefont {Y.}~\bibnamefont {Xu}}, \bibinfo {author} {\bibfnamefont
  {B.}~\bibnamefont {Bradlyn}},\ and\ \bibinfo {author} {\bibfnamefont {B.~A.}\
  \bibnamefont {Bernevig}},\ }\bibfield  {title} {\bibinfo {title} {{Magnetic
  topological quantum chemistry}},\ }\href
  {https://doi.org/10.1038/s41467-021-26241-8} {\bibfield  {journal} {\bibinfo
  {journal} {Nat. Commun.}\ }\textbf {\bibinfo {volume} {12}},\ \bibinfo
  {pages} {5965} (\bibinfo {year} {2021})}\BibitemShut {NoStop}%
\bibitem [{\citenamefont {Bernevig}\ \emph {et~al.}(2022)\citenamefont
  {Bernevig}, \citenamefont {Felser},\ and\ \citenamefont
  {Beidenkopf}}]{bernevig-progress-2022}%
  \BibitemOpen
  \bibfield  {author} {\bibinfo {author} {\bibfnamefont {B.~A.}\ \bibnamefont
  {Bernevig}}, \bibinfo {author} {\bibfnamefont {C.}~\bibnamefont {Felser}},\
  and\ \bibinfo {author} {\bibfnamefont {H.}~\bibnamefont {Beidenkopf}},\
  }\bibfield  {title} {\bibinfo {title} {{Progress and prospects in magnetic
  topological materials}},\ }\href {https://doi.org/10.1038/s41586-021-04105-x}
  {\bibfield  {journal} {\bibinfo  {journal} {Nature}\ }\textbf {\bibinfo
  {volume} {603}},\ \bibinfo {pages} {41} (\bibinfo {year} {2022})}\BibitemShut
  {NoStop}%
\bibitem [{\citenamefont {Sun}\ \emph {et~al.}(2019)\citenamefont {Sun},
  \citenamefont {Xia}, \citenamefont {Chen}, \citenamefont {Zhang},
  \citenamefont {Liu}, \citenamefont {Yao}, \citenamefont {Tang}, \citenamefont
  {Zhao}, \citenamefont {Xu},\ and\ \citenamefont
  {Liu}}]{PhysRevLett.123.096401}%
  \BibitemOpen
  \bibfield  {author} {\bibinfo {author} {\bibfnamefont {H.}~\bibnamefont
  {Sun}}, \bibinfo {author} {\bibfnamefont {B.}~\bibnamefont {Xia}}, \bibinfo
  {author} {\bibfnamefont {Z.}~\bibnamefont {Chen}}, \bibinfo {author}
  {\bibfnamefont {Y.}~\bibnamefont {Zhang}}, \bibinfo {author} {\bibfnamefont
  {P.}~\bibnamefont {Liu}}, \bibinfo {author} {\bibfnamefont {Q.}~\bibnamefont
  {Yao}}, \bibinfo {author} {\bibfnamefont {H.}~\bibnamefont {Tang}}, \bibinfo
  {author} {\bibfnamefont {Y.}~\bibnamefont {Zhao}}, \bibinfo {author}
  {\bibfnamefont {H.}~\bibnamefont {Xu}},\ and\ \bibinfo {author}
  {\bibfnamefont {Q.}~\bibnamefont {Liu}},\ }\bibfield  {title} {\bibinfo
  {title} {{Rational Design Principles of the Quantum Anomalous Hall Effect in
  Superlatticelike Magnetic Topological Insulators}},\ }\href
  {https://doi.org/10.1103/PhysRevLett.123.096401} {\bibfield  {journal}
  {\bibinfo  {journal} {Phys. Rev. Lett.}\ }\textbf {\bibinfo {volume} {123}},\
  \bibinfo {pages} {096401} (\bibinfo {year} {2019})}\BibitemShut {NoStop}%
\bibitem [{\citenamefont {Deng}\ \emph {et~al.}(2020)\citenamefont {Deng},
  \citenamefont {Yu}, \citenamefont {Shi}, \citenamefont {Guo}, \citenamefont
  {Xu}, \citenamefont {Wang}, \citenamefont {Chen},\ and\ \citenamefont
  {Zhang}}]{doi:10.1126/science.aax8156}%
  \BibitemOpen
  \bibfield  {author} {\bibinfo {author} {\bibfnamefont {Y.}~\bibnamefont
  {Deng}}, \bibinfo {author} {\bibfnamefont {Y.}~\bibnamefont {Yu}}, \bibinfo
  {author} {\bibfnamefont {M.~Z.}\ \bibnamefont {Shi}}, \bibinfo {author}
  {\bibfnamefont {Z.}~\bibnamefont {Guo}}, \bibinfo {author} {\bibfnamefont
  {Z.}~\bibnamefont {Xu}}, \bibinfo {author} {\bibfnamefont {J.}~\bibnamefont
  {Wang}}, \bibinfo {author} {\bibfnamefont {X.~H.}\ \bibnamefont {Chen}},\
  and\ \bibinfo {author} {\bibfnamefont {Y.}~\bibnamefont {Zhang}},\ }\bibfield
   {title} {\bibinfo {title} {{Quantum anomalous Hall effect in intrinsic
  magnetic topological insulator MnBi$_2$Te$_4$}},\ }\href
  {https://doi.org/10.1126/science.aax8156} {\bibfield  {journal} {\bibinfo
  {journal} {Science}\ }\textbf {\bibinfo {volume} {367}},\ \bibinfo {pages}
  {895} (\bibinfo {year} {2020})}\BibitemShut {NoStop}%
\bibitem [{\citenamefont {Li}\ \emph {et~al.}(2021)\citenamefont {Li},
  \citenamefont {Chen}, \citenamefont {Jiang},\ and\ \citenamefont
  {Xie}}]{PhysRevLett.127.236402}%
  \BibitemOpen
  \bibfield  {author} {\bibinfo {author} {\bibfnamefont {H.}~\bibnamefont
  {Li}}, \bibinfo {author} {\bibfnamefont {C.-Z.}\ \bibnamefont {Chen}},
  \bibinfo {author} {\bibfnamefont {H.}~\bibnamefont {Jiang}},\ and\ \bibinfo
  {author} {\bibfnamefont {X.~C.}\ \bibnamefont {Xie}},\ }\bibfield  {title}
  {\bibinfo {title} {{Coexistence of Quantum Hall and Quantum Anomalous Hall
  Phases in Disordered ${\mathrm{MnBi}}_{2}{\mathrm{Te}}_{4}$}},\ }\href
  {https://doi.org/10.1103/PhysRevLett.127.236402} {\bibfield  {journal}
  {\bibinfo  {journal} {Phys. Rev. Lett.}\ }\textbf {\bibinfo {volume} {127}},\
  \bibinfo {pages} {236402} (\bibinfo {year} {2021})}\BibitemShut {NoStop}%
\bibitem [{\citenamefont {Chen}\ \emph {et~al.}(2014)\citenamefont {Chen},
  \citenamefont {Niu},\ and\ \citenamefont
  {MacDonald}}]{PhysRevLett.112.017205}%
  \BibitemOpen
  \bibfield  {author} {\bibinfo {author} {\bibfnamefont {H.}~\bibnamefont
  {Chen}}, \bibinfo {author} {\bibfnamefont {Q.}~\bibnamefont {Niu}},\ and\
  \bibinfo {author} {\bibfnamefont {A.~H.}\ \bibnamefont {MacDonald}},\
  }\bibfield  {title} {\bibinfo {title} {{Anomalous Hall Effect Arising from
  Noncollinear Antiferromagnetism}},\ }\href
  {https://doi.org/10.1103/PhysRevLett.112.017205} {\bibfield  {journal}
  {\bibinfo  {journal} {Phys. Rev. Lett.}\ }\textbf {\bibinfo {volume} {112}},\
  \bibinfo {pages} {017205} (\bibinfo {year} {2014})}\BibitemShut {NoStop}%
\bibitem [{\citenamefont {Zhang}\ \emph {et~al.}(2017)\citenamefont {Zhang},
  \citenamefont {Sun}, \citenamefont {Yang}, \citenamefont
  {\ifmmode~\check{Z}\else \v{Z}\fi{}elezn\'y}, \citenamefont {Parkin},
  \citenamefont {Felser},\ and\ \citenamefont {Yan}}]{PhysRevB.95.075128}%
  \BibitemOpen
  \bibfield  {author} {\bibinfo {author} {\bibfnamefont {Y.}~\bibnamefont
  {Zhang}}, \bibinfo {author} {\bibfnamefont {Y.}~\bibnamefont {Sun}}, \bibinfo
  {author} {\bibfnamefont {H.}~\bibnamefont {Yang}}, \bibinfo {author}
  {\bibfnamefont {J.}~\bibnamefont {\ifmmode~\check{Z}\else
  \v{Z}\fi{}elezn\'y}}, \bibinfo {author} {\bibfnamefont {S.~P.~P.}\
  \bibnamefont {Parkin}}, \bibinfo {author} {\bibfnamefont {C.}~\bibnamefont
  {Felser}},\ and\ \bibinfo {author} {\bibfnamefont {B.}~\bibnamefont {Yan}},\
  }\bibfield  {title} {\bibinfo {title} {{Strong anisotropic anomalous Hall
  effect and spin Hall effect in the chiral antiferromagnetic compounds Mn$_3$X
  ( X = Ge, Sn, Ga, Ir, Rh, and Pt)}},\ }\href
  {https://doi.org/10.1103/PhysRevB.95.075128} {\bibfield  {journal} {\bibinfo
  {journal} {Phys. Rev. B}\ }\textbf {\bibinfo {volume} {95}},\ \bibinfo
  {pages} {075128} (\bibinfo {year} {2017})}\BibitemShut {NoStop}%
\bibitem [{\citenamefont {Li}\ \emph {et~al.}(2020)\citenamefont {Li},
  \citenamefont {Koo}, \citenamefont {Ning}, \citenamefont {Li}, \citenamefont
  {Miao}, \citenamefont {Min}, \citenamefont {Zhu}, \citenamefont {Wang},
  \citenamefont {Alem}, \citenamefont {Liu}, \citenamefont {Mao},\ and\
  \citenamefont {Yan}}]{li-giant-2020}%
  \BibitemOpen
  \bibfield  {author} {\bibinfo {author} {\bibfnamefont {P.}~\bibnamefont
  {Li}}, \bibinfo {author} {\bibfnamefont {J.}~\bibnamefont {Koo}}, \bibinfo
  {author} {\bibfnamefont {W.}~\bibnamefont {Ning}}, \bibinfo {author}
  {\bibfnamefont {J.}~\bibnamefont {Li}}, \bibinfo {author} {\bibfnamefont
  {L.}~\bibnamefont {Miao}}, \bibinfo {author} {\bibfnamefont {L.}~\bibnamefont
  {Min}}, \bibinfo {author} {\bibfnamefont {Y.}~\bibnamefont {Zhu}}, \bibinfo
  {author} {\bibfnamefont {Y.}~\bibnamefont {Wang}}, \bibinfo {author}
  {\bibfnamefont {N.}~\bibnamefont {Alem}}, \bibinfo {author} {\bibfnamefont
  {C.-X.}\ \bibnamefont {Liu}}, \bibinfo {author} {\bibfnamefont
  {Z.}~\bibnamefont {Mao}},\ and\ \bibinfo {author} {\bibfnamefont
  {B.}~\bibnamefont {Yan}},\ }\bibfield  {title} {\bibinfo {title} {{Giant room
  temperature anomalous {Hall} effect and tunable topology in a ferromagnetic
  topological semimetal {Co$_2$MnAl}}},\ }\href
  {https://doi.org/10.1038/s41467-020-17174-9} {\bibfield  {journal} {\bibinfo
  {journal} {Nat. Commun.}\ }\textbf {\bibinfo {volume} {11}},\ \bibinfo
  {pages} {3476} (\bibinfo {year} {2020})}\BibitemShut {NoStop}%
\bibitem [{\citenamefont {Zhou}\ \emph {et~al.}(2022)\citenamefont {Zhou},
  \citenamefont {Zhang}, \citenamefont {Yang}, \citenamefont {Li},
  \citenamefont {Feng}, \citenamefont {Mokrousov},\ and\ \citenamefont
  {Yao}}]{PhysRevLett.129.097201}%
  \BibitemOpen
  \bibfield  {author} {\bibinfo {author} {\bibfnamefont {X.}~\bibnamefont
  {Zhou}}, \bibinfo {author} {\bibfnamefont {R.-W.}\ \bibnamefont {Zhang}},
  \bibinfo {author} {\bibfnamefont {X.}~\bibnamefont {Yang}}, \bibinfo {author}
  {\bibfnamefont {X.-P.}\ \bibnamefont {Li}}, \bibinfo {author} {\bibfnamefont
  {W.}~\bibnamefont {Feng}}, \bibinfo {author} {\bibfnamefont {Y.}~\bibnamefont
  {Mokrousov}},\ and\ \bibinfo {author} {\bibfnamefont {Y.}~\bibnamefont
  {Yao}},\ }\bibfield  {title} {\bibinfo {title} {{Disorder- and
  Topology-Enhanced Fully Spin-Polarized Currents in Nodal Chain Spin-Gapless
  Semimetals}},\ }\href {https://doi.org/10.1103/PhysRevLett.129.097201}
  {\bibfield  {journal} {\bibinfo  {journal} {Phys. Rev. Lett.}\ }\textbf
  {\bibinfo {volume} {129}},\ \bibinfo {pages} {097201} (\bibinfo {year}
  {2022})}\BibitemShut {NoStop}%
\bibitem [{\citenamefont {Guin}\ \emph
  {et~al.}(2019{\natexlab{a}})\citenamefont {Guin}, \citenamefont {Vir},
  \citenamefont {Zhang}, \citenamefont {Kumar}, \citenamefont {Watzman},
  \citenamefont {Fu}, \citenamefont {Liu}, \citenamefont {Manna}, \citenamefont
  {Schnelle}, \citenamefont {Gooth} \emph {et~al.}}]{guin2019zero}%
  \BibitemOpen
  \bibfield  {author} {\bibinfo {author} {\bibfnamefont {S.~N.}\ \bibnamefont
  {Guin}}, \bibinfo {author} {\bibfnamefont {P.}~\bibnamefont {Vir}}, \bibinfo
  {author} {\bibfnamefont {Y.}~\bibnamefont {Zhang}}, \bibinfo {author}
  {\bibfnamefont {N.}~\bibnamefont {Kumar}}, \bibinfo {author} {\bibfnamefont
  {S.~J.}\ \bibnamefont {Watzman}}, \bibinfo {author} {\bibfnamefont
  {C.}~\bibnamefont {Fu}}, \bibinfo {author} {\bibfnamefont {E.}~\bibnamefont
  {Liu}}, \bibinfo {author} {\bibfnamefont {K.}~\bibnamefont {Manna}}, \bibinfo
  {author} {\bibfnamefont {W.}~\bibnamefont {Schnelle}}, \bibinfo {author}
  {\bibfnamefont {J.}~\bibnamefont {Gooth}}, \emph {et~al.},\ }\bibfield
  {title} {\bibinfo {title} {{Zero-Field Nernst Effect in a Ferromagnetic
  Kagome-Lattice Weyl-Semimetal Co$_3$Sn$_2$S$_2$}},\ }\href
  {https://doi.org/10.1002/adma.201806622} {\bibfield  {journal} {\bibinfo
  {journal} {Adv. Mater.}\ }\textbf {\bibinfo {volume} {31}},\ \bibinfo {pages}
  {1806622} (\bibinfo {year} {2019}{\natexlab{a}})}\BibitemShut {NoStop}%
\bibitem [{\citenamefont {Guin}\ \emph
  {et~al.}(2019{\natexlab{b}})\citenamefont {Guin}, \citenamefont {Manna},
  \citenamefont {Noky}, \citenamefont {Watzman}, \citenamefont {Fu},
  \citenamefont {Kumar}, \citenamefont {Schnelle}, \citenamefont {Shekhar},
  \citenamefont {Sun}, \citenamefont {Gooth} \emph
  {et~al.}}]{guin2019anomalous}%
  \BibitemOpen
  \bibfield  {author} {\bibinfo {author} {\bibfnamefont {S.~N.}\ \bibnamefont
  {Guin}}, \bibinfo {author} {\bibfnamefont {K.}~\bibnamefont {Manna}},
  \bibinfo {author} {\bibfnamefont {J.}~\bibnamefont {Noky}}, \bibinfo {author}
  {\bibfnamefont {S.~J.}\ \bibnamefont {Watzman}}, \bibinfo {author}
  {\bibfnamefont {C.}~\bibnamefont {Fu}}, \bibinfo {author} {\bibfnamefont
  {N.}~\bibnamefont {Kumar}}, \bibinfo {author} {\bibfnamefont
  {W.}~\bibnamefont {Schnelle}}, \bibinfo {author} {\bibfnamefont
  {C.}~\bibnamefont {Shekhar}}, \bibinfo {author} {\bibfnamefont
  {Y.}~\bibnamefont {Sun}}, \bibinfo {author} {\bibfnamefont {J.}~\bibnamefont
  {Gooth}}, \emph {et~al.},\ }\bibfield  {title} {\bibinfo {title} {{Anomalous
  Nernst effect beyond the magnetization scaling relation in the ferromagnetic
  Heusler compound Co$_2$MnGa}},\ }\href
  {https://doi.org/10.1038/s41427-019-0116-z} {\bibfield  {journal} {\bibinfo
  {journal} {NPG Asia Mater.}\ }\textbf {\bibinfo {volume} {11}},\ \bibinfo
  {pages} {16} (\bibinfo {year} {2019}{\natexlab{b}})}\BibitemShut {NoStop}%
\bibitem [{\citenamefont {Zhang}\ \emph
  {et~al.}(2021{\natexlab{a}})\citenamefont {Zhang}, \citenamefont {Xu},\ and\
  \citenamefont {Ke}}]{PhysRevB.103.L201101}%
  \BibitemOpen
  \bibfield  {author} {\bibinfo {author} {\bibfnamefont {H.}~\bibnamefont
  {Zhang}}, \bibinfo {author} {\bibfnamefont {C.~Q.}\ \bibnamefont {Xu}},\ and\
  \bibinfo {author} {\bibfnamefont {X.}~\bibnamefont {Ke}},\ }\bibfield
  {title} {\bibinfo {title} {{Topological Nernst effect, anomalous Nernst
  effect, and anomalous thermal Hall effect in the Dirac semimetal
  ${\mathrm{Fe}}_{3}{\mathrm{Sn}}_{2}$}},\ }\href
  {https://doi.org/10.1103/PhysRevB.103.L201101} {\bibfield  {journal}
  {\bibinfo  {journal} {Phys. Rev. B}\ }\textbf {\bibinfo {volume} {103}},\
  \bibinfo {pages} {L201101} (\bibinfo {year}
  {2021}{\natexlab{a}})}\BibitemShut {NoStop}%
\bibitem [{\citenamefont {Pan}\ \emph {et~al.}(2022)\citenamefont {Pan},
  \citenamefont {Le}, \citenamefont {He}, \citenamefont {Watzman},
  \citenamefont {Yao}, \citenamefont {Gooth}, \citenamefont {Heremans},
  \citenamefont {Sun},\ and\ \citenamefont {Felser}}]{pan-giant-2022}%
  \BibitemOpen
  \bibfield  {author} {\bibinfo {author} {\bibfnamefont {Y.}~\bibnamefont
  {Pan}}, \bibinfo {author} {\bibfnamefont {C.}~\bibnamefont {Le}}, \bibinfo
  {author} {\bibfnamefont {B.}~\bibnamefont {He}}, \bibinfo {author}
  {\bibfnamefont {S.~J.}\ \bibnamefont {Watzman}}, \bibinfo {author}
  {\bibfnamefont {M.}~\bibnamefont {Yao}}, \bibinfo {author} {\bibfnamefont
  {J.}~\bibnamefont {Gooth}}, \bibinfo {author} {\bibfnamefont {J.~P.}\
  \bibnamefont {Heremans}}, \bibinfo {author} {\bibfnamefont {Y.}~\bibnamefont
  {Sun}},\ and\ \bibinfo {author} {\bibfnamefont {C.}~\bibnamefont {Felser}},\
  }\bibfield  {title} {\bibinfo {title} {{Giant anomalous {Nernst} signal in
  the antiferromagnet YbMnBi$_2$}},\ }\href
  {https://doi.org/10.1038/s41563-021-01149-2} {\bibfield  {journal} {\bibinfo
  {journal} {Nat. Mater.}\ }\textbf {\bibinfo {volume} {21}},\ \bibinfo {pages}
  {203} (\bibinfo {year} {2022})}\BibitemShut {NoStop}%
\bibitem [{\citenamefont {Suzuki}\ \emph {et~al.}(2019)\citenamefont {Suzuki},
  \citenamefont {Savary}, \citenamefont {Liu}, \citenamefont {Lynn},
  \citenamefont {Balents},\ and\ \citenamefont
  {Checkelsky}}]{suzuki2019singular}%
  \BibitemOpen
  \bibfield  {author} {\bibinfo {author} {\bibfnamefont {T.}~\bibnamefont
  {Suzuki}}, \bibinfo {author} {\bibfnamefont {L.}~\bibnamefont {Savary}},
  \bibinfo {author} {\bibfnamefont {J.-P.}\ \bibnamefont {Liu}}, \bibinfo
  {author} {\bibfnamefont {J.~W.}\ \bibnamefont {Lynn}}, \bibinfo {author}
  {\bibfnamefont {L.}~\bibnamefont {Balents}},\ and\ \bibinfo {author}
  {\bibfnamefont {J.~G.}\ \bibnamefont {Checkelsky}},\ }\bibfield  {title}
  {\bibinfo {title} {{Singular angular magnetoresistance in a magnetic nodal
  semimetal}},\ }\href {10.1126/science.aat0348} {\bibfield  {journal}
  {\bibinfo  {journal} {Science}\ }\textbf {\bibinfo {volume} {365}},\ \bibinfo
  {pages} {377} (\bibinfo {year} {2019})}\BibitemShut {NoStop}%
\bibitem [{\citenamefont {Zhu}\ \emph {et~al.}(2023)\citenamefont {Zhu},
  \citenamefont {Huang}, \citenamefont {Wang}, \citenamefont {Graf},
  \citenamefont {Lin}, \citenamefont {Lee}, \citenamefont {Singleton},
  \citenamefont {Min}, \citenamefont {Palmstrom}, \citenamefont {Bansil},
  \citenamefont {Singh},\ and\ \citenamefont {Mao}}]{zhu-large-2023}%
  \BibitemOpen
  \bibfield  {author} {\bibinfo {author} {\bibfnamefont {Y.}~\bibnamefont
  {Zhu}}, \bibinfo {author} {\bibfnamefont {C.-Y.}\ \bibnamefont {Huang}},
  \bibinfo {author} {\bibfnamefont {Y.}~\bibnamefont {Wang}}, \bibinfo {author}
  {\bibfnamefont {D.}~\bibnamefont {Graf}}, \bibinfo {author} {\bibfnamefont
  {H.}~\bibnamefont {Lin}}, \bibinfo {author} {\bibfnamefont {S.~H.}\
  \bibnamefont {Lee}}, \bibinfo {author} {\bibfnamefont {J.}~\bibnamefont
  {Singleton}}, \bibinfo {author} {\bibfnamefont {L.}~\bibnamefont {Min}},
  \bibinfo {author} {\bibfnamefont {J.~C.}\ \bibnamefont {Palmstrom}}, \bibinfo
  {author} {\bibfnamefont {A.}~\bibnamefont {Bansil}}, \bibinfo {author}
  {\bibfnamefont {B.}~\bibnamefont {Singh}},\ and\ \bibinfo {author}
  {\bibfnamefont {Z.}~\bibnamefont {Mao}},\ }\bibfield  {title} {\bibinfo
  {title} {{Large anomalous {Hall} effect and negative magnetoresistance in
  half-topological semimetals}},\ }\href
  {https://doi.org/10.1038/s42005-023-01469-6} {\bibfield  {journal} {\bibinfo
  {journal} {Commun. Phys.}\ }\textbf {\bibinfo {volume} {6}},\ \bibinfo
  {pages} {346} (\bibinfo {year} {2023})}\BibitemShut {NoStop}%
\bibitem [{\citenamefont {Yu}\ \emph {et~al.}(2022)\citenamefont {Yu},
  \citenamefont {Zhang}, \citenamefont {Liu}, \citenamefont {Wu}, \citenamefont
  {Li}, \citenamefont {Zhang}, \citenamefont {Yang},\ and\ \citenamefont
  {Yao}}]{yu2022encyclopedia}%
  \BibitemOpen
  \bibfield  {author} {\bibinfo {author} {\bibfnamefont {Z.-M.}\ \bibnamefont
  {Yu}}, \bibinfo {author} {\bibfnamefont {Z.}~\bibnamefont {Zhang}}, \bibinfo
  {author} {\bibfnamefont {G.-B.}\ \bibnamefont {Liu}}, \bibinfo {author}
  {\bibfnamefont {W.}~\bibnamefont {Wu}}, \bibinfo {author} {\bibfnamefont
  {X.-P.}\ \bibnamefont {Li}}, \bibinfo {author} {\bibfnamefont {R.-W.}\
  \bibnamefont {Zhang}}, \bibinfo {author} {\bibfnamefont {S.~A.}\ \bibnamefont
  {Yang}},\ and\ \bibinfo {author} {\bibfnamefont {Y.}~\bibnamefont {Yao}},\
  }\bibfield  {title} {\bibinfo {title} {{Encyclopedia of emergent particles in
  three-dimensional crystals}},\ }\href
  {https://doi.org/https://doi.org/10.1016/j.scib.2021.10.023} {\bibfield
  {journal} {\bibinfo  {journal} {Sci. Bull.}\ }\textbf {\bibinfo {volume}
  {67}},\ \bibinfo {pages} {375} (\bibinfo {year} {2022})}\BibitemShut
  {NoStop}%
\bibitem [{\citenamefont {Zhang}\ \emph
  {et~al.}(2022{\natexlab{a}})\citenamefont {Zhang}, \citenamefont {Liu},
  \citenamefont {Yu}, \citenamefont {Yang},\ and\ \citenamefont
  {Yao}}]{PhysRevB.105.104426}%
  \BibitemOpen
  \bibfield  {author} {\bibinfo {author} {\bibfnamefont {Z.}~\bibnamefont
  {Zhang}}, \bibinfo {author} {\bibfnamefont {G.-B.}\ \bibnamefont {Liu}},
  \bibinfo {author} {\bibfnamefont {Z.-M.}\ \bibnamefont {Yu}}, \bibinfo
  {author} {\bibfnamefont {S.~A.}\ \bibnamefont {Yang}},\ and\ \bibinfo
  {author} {\bibfnamefont {Y.}~\bibnamefont {Yao}},\ }\bibfield  {title}
  {\bibinfo {title} {{Encyclopedia of emergent particles in {type-IV} magnetic
  space groups}},\ }\href {https://doi.org/10.1103/PhysRevB.105.104426}
  {\bibfield  {journal} {\bibinfo  {journal} {Phys. Rev. B}\ }\textbf {\bibinfo
  {volume} {105}},\ \bibinfo {pages} {104426} (\bibinfo {year}
  {2022}{\natexlab{a}})}\BibitemShut {NoStop}%
\bibitem [{\citenamefont {Liu}\ \emph {et~al.}(2022{\natexlab{a}})\citenamefont
  {Liu}, \citenamefont {Zhang}, \citenamefont {Yu}, \citenamefont {Yang},\ and\
  \citenamefont {Yao}}]{PhysRevB.105.085117}%
  \BibitemOpen
  \bibfield  {author} {\bibinfo {author} {\bibfnamefont {G.-B.}\ \bibnamefont
  {Liu}}, \bibinfo {author} {\bibfnamefont {Z.}~\bibnamefont {Zhang}}, \bibinfo
  {author} {\bibfnamefont {Z.-M.}\ \bibnamefont {Yu}}, \bibinfo {author}
  {\bibfnamefont {S.~A.}\ \bibnamefont {Yang}},\ and\ \bibinfo {author}
  {\bibfnamefont {Y.}~\bibnamefont {Yao}},\ }\bibfield  {title} {\bibinfo
  {title} {{Systematic investigation of emergent particles in {type-III}
  magnetic space groups}},\ }\href
  {https://doi.org/10.1103/PhysRevB.105.085117} {\bibfield  {journal} {\bibinfo
   {journal} {Phys. Rev. B}\ }\textbf {\bibinfo {volume} {105}},\ \bibinfo
  {pages} {085117} (\bibinfo {year} {2022}{\natexlab{a}})}\BibitemShut
  {NoStop}%
\bibitem [{\citenamefont {Zhang}\ \emph
  {et~al.}(2021{\natexlab{b}})\citenamefont {Zhang}, \citenamefont {Yu},\ and\
  \citenamefont {Yang}}]{PhysRevB.103.115112}%
  \BibitemOpen
  \bibfield  {author} {\bibinfo {author} {\bibfnamefont {Z.}~\bibnamefont
  {Zhang}}, \bibinfo {author} {\bibfnamefont {Z.-M.}\ \bibnamefont {Yu}},\ and\
  \bibinfo {author} {\bibfnamefont {S.~A.}\ \bibnamefont {Yang}},\ }\bibfield
  {title} {\bibinfo {title} {{Magnetic higher-order nodal lines}},\ }\href
  {https://doi.org/10.1103/PhysRevB.103.115112} {\bibfield  {journal} {\bibinfo
   {journal} {Phys. Rev. B}\ }\textbf {\bibinfo {volume} {103}},\ \bibinfo
  {pages} {115112} (\bibinfo {year} {2021}{\natexlab{b}})}\BibitemShut
  {NoStop}%
\bibitem [{\citenamefont {Zhang}\ \emph {et~al.}(2024)\citenamefont {Zhang},
  \citenamefont {Cui}, \citenamefont {Li}, \citenamefont {Duan}, \citenamefont
  {Li}, \citenamefont {Yu},\ and\ \citenamefont {Yao}}]{zhang2023predictable}%
  \BibitemOpen
  \bibfield  {author} {\bibinfo {author} {\bibfnamefont {R.-W.}\ \bibnamefont
  {Zhang}}, \bibinfo {author} {\bibfnamefont {C.}~\bibnamefont {Cui}}, \bibinfo
  {author} {\bibfnamefont {R.}~\bibnamefont {Li}}, \bibinfo {author}
  {\bibfnamefont {J.}~\bibnamefont {Duan}}, \bibinfo {author} {\bibfnamefont
  {L.}~\bibnamefont {Li}}, \bibinfo {author} {\bibfnamefont {Z.-M.}\
  \bibnamefont {Yu}},\ and\ \bibinfo {author} {\bibfnamefont {Y.}~\bibnamefont
  {Yao}},\ }\bibfield  {title} {\bibinfo {title} {{Predictable Gate-Field
  Control of Spin in Altermagnets with Spin-Layer Coupling}},\ }\href
  {https://doi.org/10.1103/PhysRevLett.133.056401} {\bibfield  {journal}
  {\bibinfo  {journal} {Phys. Rev. Lett.}\ }\textbf {\bibinfo {volume} {133}},\
  \bibinfo {pages} {056401} (\bibinfo {year} {2024})}\BibitemShut {NoStop}%
\bibitem [{\citenamefont {Ma}\ \emph {et~al.}(2021)\citenamefont {Ma},
  \citenamefont {Hu}, \citenamefont {Li}, \citenamefont {Liu}, \citenamefont
  {Yao}, \citenamefont {Jia},\ and\ \citenamefont
  {Liu}}]{ma2021multifunctional}%
  \BibitemOpen
  \bibfield  {author} {\bibinfo {author} {\bibfnamefont {H.-Y.}\ \bibnamefont
  {Ma}}, \bibinfo {author} {\bibfnamefont {M.}~\bibnamefont {Hu}}, \bibinfo
  {author} {\bibfnamefont {N.}~\bibnamefont {Li}}, \bibinfo {author}
  {\bibfnamefont {J.}~\bibnamefont {Liu}}, \bibinfo {author} {\bibfnamefont
  {W.}~\bibnamefont {Yao}}, \bibinfo {author} {\bibfnamefont {J.-F.}\
  \bibnamefont {Jia}},\ and\ \bibinfo {author} {\bibfnamefont {J.}~\bibnamefont
  {Liu}},\ }\bibfield  {title} {\bibinfo {title} {Multifunctional
  antiferromagnetic materials with giant piezomagnetism and noncollinear spin
  current},\ }\href {https://doi.org/10.1038/s41467-021-23127-7} {\bibfield
  {journal} {\bibinfo  {journal} {Nat. Commun.}\ }\textbf {\bibinfo {volume}
  {12}},\ \bibinfo {pages} {2846} (\bibinfo {year} {2021})}\BibitemShut
  {NoStop}%
\bibitem [{\citenamefont {Han}\ \emph {et~al.}(2024)\citenamefont {Han},
  \citenamefont {Fu}, \citenamefont {Peng}, \citenamefont {Cheng},
  \citenamefont {Dai}, \citenamefont {Liu}, \citenamefont {Li}, \citenamefont
  {Zhang}, \citenamefont {Zhu}, \citenamefont {Bai}, \citenamefont {Zhou},
  \citenamefont {Liang}, \citenamefont {Chen}, \citenamefont {Wang},
  \citenamefont {Chen}, \citenamefont {Yang}, \citenamefont {Zhang},
  \citenamefont {Song}, \citenamefont {Liu},\ and\ \citenamefont
  {Pan}}]{doi:10.1126/sciadv.adn0479}%
  \BibitemOpen
  \bibfield  {author} {\bibinfo {author} {\bibfnamefont {L.}~\bibnamefont
  {Han}}, \bibinfo {author} {\bibfnamefont {X.}~\bibnamefont {Fu}}, \bibinfo
  {author} {\bibfnamefont {R.}~\bibnamefont {Peng}}, \bibinfo {author}
  {\bibfnamefont {X.}~\bibnamefont {Cheng}}, \bibinfo {author} {\bibfnamefont
  {J.}~\bibnamefont {Dai}}, \bibinfo {author} {\bibfnamefont {L.}~\bibnamefont
  {Liu}}, \bibinfo {author} {\bibfnamefont {Y.}~\bibnamefont {Li}}, \bibinfo
  {author} {\bibfnamefont {Y.}~\bibnamefont {Zhang}}, \bibinfo {author}
  {\bibfnamefont {W.}~\bibnamefont {Zhu}}, \bibinfo {author} {\bibfnamefont
  {H.}~\bibnamefont {Bai}}, \bibinfo {author} {\bibfnamefont {Y.}~\bibnamefont
  {Zhou}}, \bibinfo {author} {\bibfnamefont {S.}~\bibnamefont {Liang}},
  \bibinfo {author} {\bibfnamefont {C.}~\bibnamefont {Chen}}, \bibinfo {author}
  {\bibfnamefont {Q.}~\bibnamefont {Wang}}, \bibinfo {author} {\bibfnamefont
  {X.}~\bibnamefont {Chen}}, \bibinfo {author} {\bibfnamefont {L.}~\bibnamefont
  {Yang}}, \bibinfo {author} {\bibfnamefont {Y.}~\bibnamefont {Zhang}},
  \bibinfo {author} {\bibfnamefont {C.}~\bibnamefont {Song}}, \bibinfo {author}
  {\bibfnamefont {J.}~\bibnamefont {Liu}},\ and\ \bibinfo {author}
  {\bibfnamefont {F.}~\bibnamefont {Pan}},\ }\bibfield  {title} {\bibinfo
  {title} {{Electrical 180° switching of Néel vector in spin-splitting
  antiferromagnet}},\ }\href {https://doi.org/10.1126/sciadv.adn0479}
  {\bibfield  {journal} {\bibinfo  {journal} {Sci. Adv.}\ }\textbf {\bibinfo
  {volume} {10}},\ \bibinfo {pages} {eadn0479} (\bibinfo {year}
  {2024})}\BibitemShut {NoStop}%
\bibitem [{\citenamefont {Shao}\ and\ \citenamefont
  {Tsymbal}(2024)}]{shaoantiferromagnetic2024}%
  \BibitemOpen
  \bibfield  {author} {\bibinfo {author} {\bibfnamefont {D.-F.}\ \bibnamefont
  {Shao}}\ and\ \bibinfo {author} {\bibfnamefont {E.~Y.}\ \bibnamefont
  {Tsymbal}},\ }\bibfield  {title} {\bibinfo {title} {Antiferromagnetic tunnel
  junctions for spintronics},\ }\href
  {https://doi.org/10.1038/s44306-024-00014-7} {\bibfield  {journal} {\bibinfo
  {journal} {npj Spintronics}\ }\textbf {\bibinfo {volume} {2}},\ \bibinfo
  {pages} {13} (\bibinfo {year} {2024})}\BibitemShut {NoStop}%
\bibitem [{\citenamefont {Feng}\ \emph {et~al.}(2019)\citenamefont {Feng},
  \citenamefont {Yan},\ and\ \citenamefont {Liu}}]{aelm.201800466}%
  \BibitemOpen
  \bibfield  {author} {\bibinfo {author} {\bibfnamefont {Z.}~\bibnamefont
  {Feng}}, \bibinfo {author} {\bibfnamefont {H.}~\bibnamefont {Yan}},\ and\
  \bibinfo {author} {\bibfnamefont {Z.}~\bibnamefont {Liu}},\ }\bibfield
  {title} {\bibinfo {title} {{Electric-Field Control of Magnetic Order: From
  FeRh to Topological Antiferromagnetic Spintronics}},\ }\href
  {https://doi.org/https://doi.org/10.1002/aelm.201800466} {\bibfield
  {journal} {\bibinfo  {journal} {Adv. Electron. Mater.}\ }\textbf {\bibinfo
  {volume} {5}},\ \bibinfo {pages} {1800466} (\bibinfo {year}
  {2019})}\BibitemShut {NoStop}%
\bibitem [{\citenamefont {Tsai}\ \emph {et~al.}(2020)\citenamefont {Tsai},
  \citenamefont {Higo}, \citenamefont {Kondou}, \citenamefont {Nomoto},
  \citenamefont {Sakai}, \citenamefont {Kobayashi}, \citenamefont {Nakano},
  \citenamefont {Yakushiji}, \citenamefont {Arita}, \citenamefont {Miwa},
  \citenamefont {Otani},\ and\ \citenamefont
  {Nakatsuji}}]{WOS:000527521300001}%
  \BibitemOpen
  \bibfield  {author} {\bibinfo {author} {\bibfnamefont {H.}~\bibnamefont
  {Tsai}}, \bibinfo {author} {\bibfnamefont {T.}~\bibnamefont {Higo}}, \bibinfo
  {author} {\bibfnamefont {K.}~\bibnamefont {Kondou}}, \bibinfo {author}
  {\bibfnamefont {T.}~\bibnamefont {Nomoto}}, \bibinfo {author} {\bibfnamefont
  {A.}~\bibnamefont {Sakai}}, \bibinfo {author} {\bibfnamefont
  {A.}~\bibnamefont {Kobayashi}}, \bibinfo {author} {\bibfnamefont
  {T.}~\bibnamefont {Nakano}}, \bibinfo {author} {\bibfnamefont
  {K.}~\bibnamefont {Yakushiji}}, \bibinfo {author} {\bibfnamefont
  {R.}~\bibnamefont {Arita}}, \bibinfo {author} {\bibfnamefont
  {S.}~\bibnamefont {Miwa}}, \bibinfo {author} {\bibfnamefont {Y.}~\bibnamefont
  {Otani}},\ and\ \bibinfo {author} {\bibfnamefont {S.}~\bibnamefont
  {Nakatsuji}},\ }\bibfield  {title} {\bibinfo {title} {{Electrical
  manipulation of a topological antiferromagnetic state}},\ }\href
  {https://doi.org/10.1038/s41586-020-2211-2} {\bibfield  {journal} {\bibinfo
  {journal} {Nature}\ }\textbf {\bibinfo {volume} {580}},\ \bibinfo {pages}
  {608–613} (\bibinfo {year} {2020})}\BibitemShut {NoStop}%
\bibitem [{\citenamefont {Yan}\ \emph {et~al.}(2020)\citenamefont {Yan},
  \citenamefont {Feng}, \citenamefont {Qin}, \citenamefont {Zhou},
  \citenamefont {Guo}, \citenamefont {Wang}, \citenamefont {Chen},
  \citenamefont {Zhang}, \citenamefont {Wu}, \citenamefont {Jiang},\ and\
  \citenamefont {Liu}}]{yan-electricfieldcontrolled-2020}%
  \BibitemOpen
  \bibfield  {author} {\bibinfo {author} {\bibfnamefont {H.}~\bibnamefont
  {Yan}}, \bibinfo {author} {\bibfnamefont {Z.}~\bibnamefont {Feng}}, \bibinfo
  {author} {\bibfnamefont {P.}~\bibnamefont {Qin}}, \bibinfo {author}
  {\bibfnamefont {X.}~\bibnamefont {Zhou}}, \bibinfo {author} {\bibfnamefont
  {H.}~\bibnamefont {Guo}}, \bibinfo {author} {\bibfnamefont {X.}~\bibnamefont
  {Wang}}, \bibinfo {author} {\bibfnamefont {H.}~\bibnamefont {Chen}}, \bibinfo
  {author} {\bibfnamefont {X.}~\bibnamefont {Zhang}}, \bibinfo {author}
  {\bibfnamefont {H.}~\bibnamefont {Wu}}, \bibinfo {author} {\bibfnamefont
  {C.}~\bibnamefont {Jiang}},\ and\ \bibinfo {author} {\bibfnamefont
  {Z.}~\bibnamefont {Liu}},\ }\bibfield  {title} {\bibinfo {title}
  {{Electric‐{Field}‐{Controlled} {Antiferromagnetic} {Spintronic}
  {Devices}}},\ }\href {https://doi.org/10.1002/adma.201905603} {\bibfield
  {journal} {\bibinfo  {journal} {Adv. Mater.}\ }\textbf {\bibinfo {volume}
  {32}},\ \bibinfo {pages} {1905603} (\bibinfo {year} {2020})}\BibitemShut
  {NoStop}%
\bibitem [{\citenamefont {Hu}\ \emph {et~al.}(2022)\citenamefont {Hu},
  \citenamefont {Shao}, \citenamefont {Yang}, \citenamefont {Pan},
  \citenamefont {Fu}, \citenamefont {Tang}, \citenamefont {Yang}, \citenamefont
  {Fan}, \citenamefont {Zhou}, \citenamefont {Tsymbal} \emph
  {et~al.}}]{hu2022efficient}%
  \BibitemOpen
  \bibfield  {author} {\bibinfo {author} {\bibfnamefont {S.}~\bibnamefont
  {Hu}}, \bibinfo {author} {\bibfnamefont {D.-F.}\ \bibnamefont {Shao}},
  \bibinfo {author} {\bibfnamefont {H.}~\bibnamefont {Yang}}, \bibinfo {author}
  {\bibfnamefont {C.}~\bibnamefont {Pan}}, \bibinfo {author} {\bibfnamefont
  {Z.}~\bibnamefont {Fu}}, \bibinfo {author} {\bibfnamefont {M.}~\bibnamefont
  {Tang}}, \bibinfo {author} {\bibfnamefont {Y.}~\bibnamefont {Yang}}, \bibinfo
  {author} {\bibfnamefont {W.}~\bibnamefont {Fan}}, \bibinfo {author}
  {\bibfnamefont {S.}~\bibnamefont {Zhou}}, \bibinfo {author} {\bibfnamefont
  {E.~Y.}\ \bibnamefont {Tsymbal}}, \emph {et~al.},\ }\bibfield  {title}
  {\bibinfo {title} {{Efficient perpendicular magnetization switching by a
  magnetic spin Hall effect in a noncollinear antiferromagnet}},\ }\href
  {https://doi.org/10.1038/s41467-022-32179-2} {\bibfield  {journal} {\bibinfo
  {journal} {Nat. Commun.}\ }\textbf {\bibinfo {volume} {13}},\ \bibinfo
  {pages} {4447} (\bibinfo {year} {2022})}\BibitemShut {NoStop}%
\bibitem [{\citenamefont {Kim}\ \emph {et~al.}(2022)\citenamefont {Kim},
  \citenamefont {Beach}, \citenamefont {Lee}, \citenamefont {Ono},
  \citenamefont {Rasing},\ and\ \citenamefont {Yang}}]{kim2022ferrimagnetic}%
  \BibitemOpen
  \bibfield  {author} {\bibinfo {author} {\bibfnamefont {S.~K.}\ \bibnamefont
  {Kim}}, \bibinfo {author} {\bibfnamefont {G.~S.}\ \bibnamefont {Beach}},
  \bibinfo {author} {\bibfnamefont {K.-J.}\ \bibnamefont {Lee}}, \bibinfo
  {author} {\bibfnamefont {T.}~\bibnamefont {Ono}}, \bibinfo {author}
  {\bibfnamefont {T.}~\bibnamefont {Rasing}},\ and\ \bibinfo {author}
  {\bibfnamefont {H.}~\bibnamefont {Yang}},\ }\bibfield  {title} {\bibinfo
  {title} {{Ferrimagnetic spintronics}},\ }\href
  {https://doi.org/https://doi.org/10.1038/s41563-021-01139-4} {\bibfield
  {journal} {\bibinfo  {journal} {Nat. Mater.}\ }\textbf {\bibinfo {volume}
  {21}},\ \bibinfo {pages} {24} (\bibinfo {year} {2022})}\BibitemShut {NoStop}%
\bibitem [{\citenamefont {Chen}\ \emph {et~al.}(2024)\citenamefont {Chen},
  \citenamefont {Liu}, \citenamefont {Zhou}, \citenamefont {Meng},
  \citenamefont {Wang}, \citenamefont {Duan}, \citenamefont {Zhao},
  \citenamefont {Yan}, \citenamefont {Qin},\ and\ \citenamefont
  {Liu}}]{chen2024emerging}%
  \BibitemOpen
  \bibfield  {author} {\bibinfo {author} {\bibfnamefont {H.}~\bibnamefont
  {Chen}}, \bibinfo {author} {\bibfnamefont {L.}~\bibnamefont {Liu}}, \bibinfo
  {author} {\bibfnamefont {X.}~\bibnamefont {Zhou}}, \bibinfo {author}
  {\bibfnamefont {Z.}~\bibnamefont {Meng}}, \bibinfo {author} {\bibfnamefont
  {X.}~\bibnamefont {Wang}}, \bibinfo {author} {\bibfnamefont {Z.}~\bibnamefont
  {Duan}}, \bibinfo {author} {\bibfnamefont {G.}~\bibnamefont {Zhao}}, \bibinfo
  {author} {\bibfnamefont {H.}~\bibnamefont {Yan}}, \bibinfo {author}
  {\bibfnamefont {P.}~\bibnamefont {Qin}},\ and\ \bibinfo {author}
  {\bibfnamefont {Z.}~\bibnamefont {Liu}},\ }\bibfield  {title} {\bibinfo
  {title} {{Emerging Antiferromagnets for Spintronics}},\ }\href
  {https://doi.org/https://doi.org/10.1002/adma.202310379} {\bibfield
  {journal} {\bibinfo  {journal} {Adv. Mater.}\ }\textbf {\bibinfo {volume}
  {36}},\ \bibinfo {pages} {2310379} (\bibinfo {year} {2024})}\BibitemShut
  {NoStop}%
\bibitem [{\citenamefont {Gong}\ \emph {et~al.}(2024)\citenamefont {Gong},
  \citenamefont {Wang}, \citenamefont {Han}, \citenamefont {Cheng},
  \citenamefont {Wang}, \citenamefont {Yu},\ and\ \citenamefont
  {Yao}}]{gong2023hidden}%
  \BibitemOpen
  \bibfield  {author} {\bibinfo {author} {\bibfnamefont {J.}~\bibnamefont
  {Gong}}, \bibinfo {author} {\bibfnamefont {Y.}~\bibnamefont {Wang}}, \bibinfo
  {author} {\bibfnamefont {Y.}~\bibnamefont {Han}}, \bibinfo {author}
  {\bibfnamefont {Z.}~\bibnamefont {Cheng}}, \bibinfo {author} {\bibfnamefont
  {X.}~\bibnamefont {Wang}}, \bibinfo {author} {\bibfnamefont {Z.-M.}\
  \bibnamefont {Yu}},\ and\ \bibinfo {author} {\bibfnamefont {Y.}~\bibnamefont
  {Yao}},\ }\bibfield  {title} {\bibinfo {title} {{Hidden Real Topology and
  Unusual Magnetoelectric Responses in Two-Dimensional Antiferromagnets}},\
  }\href {https://doi.org/https://doi.org/10.1002/adma.202402232} {\bibfield
  {journal} {\bibinfo  {journal} {Adv. Mater.}\ }\textbf {\bibinfo {volume}
  {36}},\ \bibinfo {pages} {2402232} (\bibinfo {year} {2024})}\BibitemShut
  {NoStop}%
\bibitem [{\citenamefont {Wang}\ \emph {et~al.}(2018)\citenamefont {Wang},
  \citenamefont {Xu}, \citenamefont {Lou}, \citenamefont {Liu}, \citenamefont
  {Li}, \citenamefont {Huang}, \citenamefont {Shen}, \citenamefont {Weng},
  \citenamefont {Wang},\ and\ \citenamefont {Lei}}]{wanglarge2018}%
  \BibitemOpen
  \bibfield  {author} {\bibinfo {author} {\bibfnamefont {Q.}~\bibnamefont
  {Wang}}, \bibinfo {author} {\bibfnamefont {Y.}~\bibnamefont {Xu}}, \bibinfo
  {author} {\bibfnamefont {R.}~\bibnamefont {Lou}}, \bibinfo {author}
  {\bibfnamefont {Z.}~\bibnamefont {Liu}}, \bibinfo {author} {\bibfnamefont
  {M.}~\bibnamefont {Li}}, \bibinfo {author} {\bibfnamefont {Y.}~\bibnamefont
  {Huang}}, \bibinfo {author} {\bibfnamefont {D.}~\bibnamefont {Shen}},
  \bibinfo {author} {\bibfnamefont {H.}~\bibnamefont {Weng}}, \bibinfo {author}
  {\bibfnamefont {S.}~\bibnamefont {Wang}},\ and\ \bibinfo {author}
  {\bibfnamefont {H.}~\bibnamefont {Lei}},\ }\bibfield  {title} {\bibinfo
  {title} {{Large intrinsic anomalous Hall effect in half-metallic ferromagnet
  Co$_3$Sn$_2$S$_2$ with magnetic Weyl fermions}},\ }\href
  {https://doi.org/10.1038/s41467-018-06088-2} {\bibfield  {journal} {\bibinfo
  {journal} {Nat. Commun.}\ }\textbf {\bibinfo {volume} {9}},\ \bibinfo {pages}
  {3681} (\bibinfo {year} {2018})}\BibitemShut {NoStop}%
\bibitem [{\citenamefont {Chang}\ \emph {et~al.}(2017)\citenamefont {Chang},
  \citenamefont {Xu}, \citenamefont {Zhou}, \citenamefont {Huang},
  \citenamefont {Singh}, \citenamefont {Wang}, \citenamefont {Belopolski},
  \citenamefont {Yin}, \citenamefont {Zhang}, \citenamefont {Bansil},
  \citenamefont {Lin},\ and\ \citenamefont {Hasan}}]{PhysRevLett.119.156401}%
  \BibitemOpen
  \bibfield  {author} {\bibinfo {author} {\bibfnamefont {G.}~\bibnamefont
  {Chang}}, \bibinfo {author} {\bibfnamefont {S.-Y.}\ \bibnamefont {Xu}},
  \bibinfo {author} {\bibfnamefont {X.}~\bibnamefont {Zhou}}, \bibinfo {author}
  {\bibfnamefont {S.-M.}\ \bibnamefont {Huang}}, \bibinfo {author}
  {\bibfnamefont {B.}~\bibnamefont {Singh}}, \bibinfo {author} {\bibfnamefont
  {B.}~\bibnamefont {Wang}}, \bibinfo {author} {\bibfnamefont {I.}~\bibnamefont
  {Belopolski}}, \bibinfo {author} {\bibfnamefont {J.}~\bibnamefont {Yin}},
  \bibinfo {author} {\bibfnamefont {S.}~\bibnamefont {Zhang}}, \bibinfo
  {author} {\bibfnamefont {A.}~\bibnamefont {Bansil}}, \bibinfo {author}
  {\bibfnamefont {H.}~\bibnamefont {Lin}},\ and\ \bibinfo {author}
  {\bibfnamefont {M.~Z.}\ \bibnamefont {Hasan}},\ }\bibfield  {title} {\bibinfo
  {title} {{Topological Hopf and Chain Link Semimetal States and Their
  Application to ${\mathrm{Co}}_{2}\mathrm{Mn}\text{G}\text{a}$}},\ }\href
  {https://doi.org/10.1103/PhysRevLett.119.156401} {\bibfield  {journal}
  {\bibinfo  {journal} {Phys. Rev. Lett.}\ }\textbf {\bibinfo {volume} {119}},\
  \bibinfo {pages} {156401} (\bibinfo {year} {2017})}\BibitemShut {NoStop}%
\bibitem [{\citenamefont {Zhang}\ \emph
  {et~al.}(2021{\natexlab{c}})\citenamefont {Zhang}, \citenamefont {Zhou},
  \citenamefont {Zhang}, \citenamefont {Ma}, \citenamefont {Yu}, \citenamefont
  {Feng},\ and\ \citenamefont {Yao}}]{doi:10.1021/acs.nanolett.1c02968}%
  \BibitemOpen
  \bibfield  {author} {\bibinfo {author} {\bibfnamefont {R.-W.}\ \bibnamefont
  {Zhang}}, \bibinfo {author} {\bibfnamefont {X.}~\bibnamefont {Zhou}},
  \bibinfo {author} {\bibfnamefont {Z.}~\bibnamefont {Zhang}}, \bibinfo
  {author} {\bibfnamefont {D.-S.}\ \bibnamefont {Ma}}, \bibinfo {author}
  {\bibfnamefont {Z.-M.}\ \bibnamefont {Yu}}, \bibinfo {author} {\bibfnamefont
  {W.}~\bibnamefont {Feng}},\ and\ \bibinfo {author} {\bibfnamefont
  {Y.}~\bibnamefont {Yao}},\ }\bibfield  {title} {\bibinfo {title} {{Weyl
  Monoloop Semi-Half-Metal and Tunable Anomalous Hall Effect}},\ }\href
  {https://doi.org/10.1021/acs.nanolett.1c02968} {\bibfield  {journal}
  {\bibinfo  {journal} {Nano Lett.}\ }\textbf {\bibinfo {volume} {21}},\
  \bibinfo {pages} {8749} (\bibinfo {year} {2021}{\natexlab{c}})}\BibitemShut
  {NoStop}%
\bibitem [{\citenamefont {Liu}\ and\ \citenamefont
  {Balents}(2017)}]{PhysRevLett.119.087202}%
  \BibitemOpen
  \bibfield  {author} {\bibinfo {author} {\bibfnamefont {J.}~\bibnamefont
  {Liu}}\ and\ \bibinfo {author} {\bibfnamefont {L.}~\bibnamefont {Balents}},\
  }\bibfield  {title} {\bibinfo {title} {{Anomalous Hall Effect and Topological
  Defects in Antiferromagnetic Weyl Semimetals:
  ${\mathrm{Mn}}_{3}\mathrm{Sn}/\mathrm{Ge}$}},\ }\href
  {https://doi.org/10.1103/PhysRevLett.119.087202} {\bibfield  {journal}
  {\bibinfo  {journal} {Phys. Rev. Lett.}\ }\textbf {\bibinfo {volume} {119}},\
  \bibinfo {pages} {087202} (\bibinfo {year} {2017})}\BibitemShut {NoStop}%
\bibitem [{\citenamefont {Nguyen}\ and\ \citenamefont
  {Yamauchi}(2023)}]{PhysRevB.107.155126}%
  \BibitemOpen
  \bibfield  {author} {\bibinfo {author} {\bibfnamefont {T.~P.~T.}\
  \bibnamefont {Nguyen}}\ and\ \bibinfo {author} {\bibfnamefont
  {K.}~\bibnamefont {Yamauchi}},\ }\bibfield  {title} {\bibinfo {title} {{Ab
  initio prediction of anomalous Hall effect in antiferromagnetic
  ${\mathrm{CaCrO}}_{3}$}},\ }\href
  {https://doi.org/10.1103/PhysRevB.107.155126} {\bibfield  {journal} {\bibinfo
   {journal} {Phys. Rev. B}\ }\textbf {\bibinfo {volume} {107}},\ \bibinfo
  {pages} {155126} (\bibinfo {year} {2023})}\BibitemShut {NoStop}%
\bibitem [{\citenamefont {Šmejkal}\ \emph {et~al.}(2020)\citenamefont
  {Šmejkal}, \citenamefont {González-Hernández}, \citenamefont {Jungwirth},\
  and\ \citenamefont {Sinova}}]{doi:10.1126/sciadv.aaz8809}%
  \BibitemOpen
  \bibfield  {author} {\bibinfo {author} {\bibfnamefont {L.}~\bibnamefont
  {Šmejkal}}, \bibinfo {author} {\bibfnamefont {R.}~\bibnamefont
  {González-Hernández}}, \bibinfo {author} {\bibfnamefont {T.}~\bibnamefont
  {Jungwirth}},\ and\ \bibinfo {author} {\bibfnamefont {J.}~\bibnamefont
  {Sinova}},\ }\bibfield  {title} {\bibinfo {title} {{Crystal time-reversal
  symmetry breaking and spontaneous Hall effect in collinear
  antiferromagnets}},\ }\href {https://doi.org/10.1126/sciadv.aaz8809}
  {\bibfield  {journal} {\bibinfo  {journal} {Sci. Adv.}\ }\textbf {\bibinfo
  {volume} {6}},\ \bibinfo {pages} {eaaz8809} (\bibinfo {year}
  {2020})}\BibitemShut {NoStop}%
\bibitem [{\citenamefont {Gonz\'alez-Hern\'andez}\ \emph
  {et~al.}(2021)\citenamefont {Gonz\'alez-Hern\'andez}, \citenamefont
  {\ifmmode~\check{S}\else \v{S}\fi{}mejkal}, \citenamefont {V\'yborn\'y},
  \citenamefont {Yahagi}, \citenamefont {Sinova}, \citenamefont {Jungwirth},\
  and\ \citenamefont {\ifmmode~\check{Z}\else
  \v{Z}\fi{}elezn\'y}}]{PhysRevLett.126.127701}%
  \BibitemOpen
  \bibfield  {author} {\bibinfo {author} {\bibfnamefont {R.}~\bibnamefont
  {Gonz\'alez-Hern\'andez}}, \bibinfo {author} {\bibfnamefont {L.}~\bibnamefont
  {\ifmmode~\check{S}\else \v{S}\fi{}mejkal}}, \bibinfo {author} {\bibfnamefont
  {K.}~\bibnamefont {V\'yborn\'y}}, \bibinfo {author} {\bibfnamefont
  {Y.}~\bibnamefont {Yahagi}}, \bibinfo {author} {\bibfnamefont
  {J.}~\bibnamefont {Sinova}}, \bibinfo {author} {\bibfnamefont {T.~c.~v.}\
  \bibnamefont {Jungwirth}},\ and\ \bibinfo {author} {\bibfnamefont
  {J.}~\bibnamefont {\ifmmode~\check{Z}\else \v{Z}\fi{}elezn\'y}},\ }\bibfield
  {title} {\bibinfo {title} {{Efficient Electrical Spin Splitter Based on
  Nonrelativistic Collinear Antiferromagnetism}},\ }\href
  {https://doi.org/10.1103/PhysRevLett.126.127701} {\bibfield  {journal}
  {\bibinfo  {journal} {Phys. Rev. Lett.}\ }\textbf {\bibinfo {volume} {126}},\
  \bibinfo {pages} {127701} (\bibinfo {year} {2021})}\BibitemShut {NoStop}%
\bibitem [{\citenamefont {\ifmmode~\check{S}\else \v{S}\fi{}mejkal}\ \emph
  {et~al.}(2022{\natexlab{a}})\citenamefont {\ifmmode~\check{S}\else
  \v{S}\fi{}mejkal}, \citenamefont {Hellenes}, \citenamefont
  {Gonz\'alez-Hern\'andez}, \citenamefont {Sinova},\ and\ \citenamefont
  {Jungwirth}}]{PhysRevX.12.011028}%
  \BibitemOpen
  \bibfield  {author} {\bibinfo {author} {\bibfnamefont {L.}~\bibnamefont
  {\ifmmode~\check{S}\else \v{S}\fi{}mejkal}}, \bibinfo {author} {\bibfnamefont
  {A.~B.}\ \bibnamefont {Hellenes}}, \bibinfo {author} {\bibfnamefont
  {R.}~\bibnamefont {Gonz\'alez-Hern\'andez}}, \bibinfo {author} {\bibfnamefont
  {J.}~\bibnamefont {Sinova}},\ and\ \bibinfo {author} {\bibfnamefont
  {T.}~\bibnamefont {Jungwirth}},\ }\bibfield  {title} {\bibinfo {title}
  {{Giant and Tunneling Magnetoresistance in Unconventional Collinear
  Antiferromagnets with Nonrelativistic Spin-Momentum Coupling}},\ }\href
  {https://doi.org/10.1103/PhysRevX.12.011028} {\bibfield  {journal} {\bibinfo
  {journal} {Phys. Rev. X}\ }\textbf {\bibinfo {volume} {12}},\ \bibinfo
  {pages} {011028} (\bibinfo {year} {2022}{\natexlab{a}})}\BibitemShut
  {NoStop}%
\bibitem [{\citenamefont {\ifmmode~\check{S}\else \v{S}\fi{}mejkal}\ \emph
  {et~al.}(2022{\natexlab{b}})\citenamefont {\ifmmode~\check{S}\else
  \v{S}\fi{}mejkal}, \citenamefont {Sinova},\ and\ \citenamefont
  {Jungwirth}}]{PhysRevX.12.031042}%
  \BibitemOpen
  \bibfield  {author} {\bibinfo {author} {\bibfnamefont {L.}~\bibnamefont
  {\ifmmode~\check{S}\else \v{S}\fi{}mejkal}}, \bibinfo {author} {\bibfnamefont
  {J.}~\bibnamefont {Sinova}},\ and\ \bibinfo {author} {\bibfnamefont
  {T.}~\bibnamefont {Jungwirth}},\ }\bibfield  {title} {\bibinfo {title}
  {{Beyond Conventional Ferromagnetism and Antiferromagnetism: A Phase with
  Nonrelativistic Spin and Crystal Rotation Symmetry}},\ }\href
  {https://doi.org/10.1103/PhysRevX.12.031042} {\bibfield  {journal} {\bibinfo
  {journal} {Phys. Rev. X}\ }\textbf {\bibinfo {volume} {12}},\ \bibinfo
  {pages} {031042} (\bibinfo {year} {2022}{\natexlab{b}})}\BibitemShut
  {NoStop}%
\bibitem [{\citenamefont {\ifmmode~\check{S}\else \v{S}\fi{}mejkal}\ \emph
  {et~al.}(2022{\natexlab{c}})\citenamefont {\ifmmode~\check{S}\else
  \v{S}\fi{}mejkal}, \citenamefont {Sinova},\ and\ \citenamefont
  {Jungwirth}}]{PhysRevX.12.040501}%
  \BibitemOpen
  \bibfield  {author} {\bibinfo {author} {\bibfnamefont {L.}~\bibnamefont
  {\ifmmode~\check{S}\else \v{S}\fi{}mejkal}}, \bibinfo {author} {\bibfnamefont
  {J.}~\bibnamefont {Sinova}},\ and\ \bibinfo {author} {\bibfnamefont
  {T.}~\bibnamefont {Jungwirth}},\ }\bibfield  {title} {\bibinfo {title}
  {{Emerging Research Landscape of Altermagnetism}},\ }\href
  {https://doi.org/10.1103/PhysRevX.12.040501} {\bibfield  {journal} {\bibinfo
  {journal} {Phys. Rev. X}\ }\textbf {\bibinfo {volume} {12}},\ \bibinfo
  {pages} {040501} (\bibinfo {year} {2022}{\natexlab{c}})}\BibitemShut
  {NoStop}%
\bibitem [{\citenamefont {Shao}\ \emph {et~al.}(2023)\citenamefont {Shao},
  \citenamefont {Jiang}, \citenamefont {Ding}, \citenamefont {Zhang},
  \citenamefont {Wang}, \citenamefont {Xiao}, \citenamefont {Gurung},
  \citenamefont {Lu}, \citenamefont {Sun},\ and\ \citenamefont
  {Tsymbal}}]{PhysRevLett.130.216702}%
  \BibitemOpen
  \bibfield  {author} {\bibinfo {author} {\bibfnamefont {D.-F.}\ \bibnamefont
  {Shao}}, \bibinfo {author} {\bibfnamefont {Y.-Y.}\ \bibnamefont {Jiang}},
  \bibinfo {author} {\bibfnamefont {J.}~\bibnamefont {Ding}}, \bibinfo {author}
  {\bibfnamefont {S.-H.}\ \bibnamefont {Zhang}}, \bibinfo {author}
  {\bibfnamefont {Z.-A.}\ \bibnamefont {Wang}}, \bibinfo {author}
  {\bibfnamefont {R.-C.}\ \bibnamefont {Xiao}}, \bibinfo {author}
  {\bibfnamefont {G.}~\bibnamefont {Gurung}}, \bibinfo {author} {\bibfnamefont
  {W.~J.}\ \bibnamefont {Lu}}, \bibinfo {author} {\bibfnamefont {Y.~P.}\
  \bibnamefont {Sun}},\ and\ \bibinfo {author} {\bibfnamefont {E.~Y.}\
  \bibnamefont {Tsymbal}},\ }\bibfield  {title} {\bibinfo {title} {{N\'eel Spin
  Currents in Antiferromagnets}},\ }\href
  {https://doi.org/10.1103/PhysRevLett.130.216702} {\bibfield  {journal}
  {\bibinfo  {journal} {Phys. Rev. Lett.}\ }\textbf {\bibinfo {volume} {130}},\
  \bibinfo {pages} {216702} (\bibinfo {year} {2023})}\BibitemShut {NoStop}%
\bibitem [{\citenamefont {Zhan}\ \emph {et~al.}(2023)\citenamefont {Zhan},
  \citenamefont {Li}, \citenamefont {Shi}, \citenamefont {Chen},\ and\
  \citenamefont {Sun}}]{PhysRevB.107.224402}%
  \BibitemOpen
  \bibfield  {author} {\bibinfo {author} {\bibfnamefont {J.}~\bibnamefont
  {Zhan}}, \bibinfo {author} {\bibfnamefont {J.}~\bibnamefont {Li}}, \bibinfo
  {author} {\bibfnamefont {W.}~\bibnamefont {Shi}}, \bibinfo {author}
  {\bibfnamefont {X.-Q.}\ \bibnamefont {Chen}},\ and\ \bibinfo {author}
  {\bibfnamefont {Y.}~\bibnamefont {Sun}},\ }\bibfield  {title} {\bibinfo
  {title} {{Coexistence of Weyl semimetal and Weyl nodal loop semimetal phases
  in a collinear antiferromagnet}},\ }\href
  {https://doi.org/10.1103/PhysRevB.107.224402} {\bibfield  {journal} {\bibinfo
   {journal} {Phys. Rev. B}\ }\textbf {\bibinfo {volume} {107}},\ \bibinfo
  {pages} {224402} (\bibinfo {year} {2023})}\BibitemShut {NoStop}%
\bibitem [{\citenamefont {Cui}\ \emph {et~al.}(2023)\citenamefont {Cui},
  \citenamefont {Zhu}, \citenamefont {Yao}, \citenamefont {Cui},\ and\
  \citenamefont {Yang}}]{PhysRevB.108.024410}%
  \BibitemOpen
  \bibfield  {author} {\bibinfo {author} {\bibfnamefont {Q.}~\bibnamefont
  {Cui}}, \bibinfo {author} {\bibfnamefont {Y.}~\bibnamefont {Zhu}}, \bibinfo
  {author} {\bibfnamefont {X.}~\bibnamefont {Yao}}, \bibinfo {author}
  {\bibfnamefont {P.}~\bibnamefont {Cui}},\ and\ \bibinfo {author}
  {\bibfnamefont {H.}~\bibnamefont {Yang}},\ }\bibfield  {title} {\bibinfo
  {title} {{Giant spin-Hall and tunneling magnetoresistance effects based on a
  two-dimensional nonrelativistic antiferromagnetic metal}},\ }\href
  {https://doi.org/10.1103/PhysRevB.108.024410} {\bibfield  {journal} {\bibinfo
   {journal} {Phys. Rev. B}\ }\textbf {\bibinfo {volume} {108}},\ \bibinfo
  {pages} {024410} (\bibinfo {year} {2023})}\BibitemShut {NoStop}%
\bibitem [{\citenamefont {Nag}\ \emph {et~al.}(2024)\citenamefont {Nag},
  \citenamefont {Das}, \citenamefont {Bhowal}, \citenamefont {Nishioka},
  \citenamefont {Bandyopadhyay}, \citenamefont {Sarker}, \citenamefont {Kumar},
  \citenamefont {Kuroda}, \citenamefont {Gopalan}, \citenamefont {Kimura},
  \citenamefont {Suresh},\ and\ \citenamefont
  {Alam}}]{nag2024gdalsiantiferromagnetictopologicalweyl}%
  \BibitemOpen
  \bibfield  {author} {\bibinfo {author} {\bibfnamefont {J.}~\bibnamefont
  {Nag}}, \bibinfo {author} {\bibfnamefont {B.}~\bibnamefont {Das}}, \bibinfo
  {author} {\bibfnamefont {S.}~\bibnamefont {Bhowal}}, \bibinfo {author}
  {\bibfnamefont {Y.}~\bibnamefont {Nishioka}}, \bibinfo {author}
  {\bibfnamefont {B.}~\bibnamefont {Bandyopadhyay}}, \bibinfo {author}
  {\bibfnamefont {S.}~\bibnamefont {Sarker}}, \bibinfo {author} {\bibfnamefont
  {S.}~\bibnamefont {Kumar}}, \bibinfo {author} {\bibfnamefont
  {K.}~\bibnamefont {Kuroda}}, \bibinfo {author} {\bibfnamefont
  {V.}~\bibnamefont {Gopalan}}, \bibinfo {author} {\bibfnamefont
  {A.}~\bibnamefont {Kimura}}, \bibinfo {author} {\bibfnamefont {K.~G.}\
  \bibnamefont {Suresh}},\ and\ \bibinfo {author} {\bibfnamefont
  {A.}~\bibnamefont {Alam}},\ }\bibfield  {title} {\bibinfo {title} {{GdAlSi:
  An antiferromagnetic topological Weyl semimetal with non-relativistic spin
  splitting}},\ }\href {https://arxiv.org/abs/2312.11980} {\bibfield  {journal}
  {\bibinfo  {journal} {arXiv:2312.11980}\ } (\bibinfo {year}
  {2024})}\BibitemShut {NoStop}%
\bibitem [{\citenamefont {Li}\ \emph {et~al.}(2017)\citenamefont {Li},
  \citenamefont {Yu}, \citenamefont {Liu}, \citenamefont {Guan}, \citenamefont
  {Wang}, \citenamefont {Zhang}, \citenamefont {Yao},\ and\ \citenamefont
  {Yang}}]{PhysRevB.96.081106}%
  \BibitemOpen
  \bibfield  {author} {\bibinfo {author} {\bibfnamefont {S.}~\bibnamefont
  {Li}}, \bibinfo {author} {\bibfnamefont {Z.-M.}\ \bibnamefont {Yu}}, \bibinfo
  {author} {\bibfnamefont {Y.}~\bibnamefont {Liu}}, \bibinfo {author}
  {\bibfnamefont {S.}~\bibnamefont {Guan}}, \bibinfo {author} {\bibfnamefont
  {S.-S.}\ \bibnamefont {Wang}}, \bibinfo {author} {\bibfnamefont
  {X.}~\bibnamefont {Zhang}}, \bibinfo {author} {\bibfnamefont
  {Y.}~\bibnamefont {Yao}},\ and\ \bibinfo {author} {\bibfnamefont {S.~A.}\
  \bibnamefont {Yang}},\ }\bibfield  {title} {\bibinfo {title} {{Type-II nodal
  loops: Theory and material realization}},\ }\href
  {https://doi.org/10.1103/PhysRevB.96.081106} {\bibfield  {journal} {\bibinfo
  {journal} {Phys. Rev. B}\ }\textbf {\bibinfo {volume} {96}},\ \bibinfo
  {pages} {081106} (\bibinfo {year} {2017})}\BibitemShut {NoStop}%
\bibitem [{\citenamefont {Yu}\ \emph {et~al.}(2019)\citenamefont {Yu},
  \citenamefont {Wu}, \citenamefont {Sheng}, \citenamefont {Zhao},\ and\
  \citenamefont {Yang}}]{PhysRevB.99.121106}%
  \BibitemOpen
  \bibfield  {author} {\bibinfo {author} {\bibfnamefont {Z.-M.}\ \bibnamefont
  {Yu}}, \bibinfo {author} {\bibfnamefont {W.}~\bibnamefont {Wu}}, \bibinfo
  {author} {\bibfnamefont {X.-L.}\ \bibnamefont {Sheng}}, \bibinfo {author}
  {\bibfnamefont {Y.~X.}\ \bibnamefont {Zhao}},\ and\ \bibinfo {author}
  {\bibfnamefont {S.~A.}\ \bibnamefont {Yang}},\ }\bibfield  {title} {\bibinfo
  {title} {{Quadratic and cubic nodal lines stabilized by crystalline
  symmetry}},\ }\href {https://doi.org/10.1103/PhysRevB.99.121106} {\bibfield
  {journal} {\bibinfo  {journal} {Phys. Rev. B}\ }\textbf {\bibinfo {volume}
  {99}},\ \bibinfo {pages} {121106} (\bibinfo {year} {2019})}\BibitemShut
  {NoStop}%
\bibitem [{\citenamefont {He}\ \emph {et~al.}(2019)\citenamefont {He},
  \citenamefont {Zhang}, \citenamefont {Meng}, \citenamefont {Jin},
  \citenamefont {Dai},\ and\ \citenamefont {Liu}}]{he2019topological}%
  \BibitemOpen
  \bibfield  {author} {\bibinfo {author} {\bibfnamefont {T.}~\bibnamefont
  {He}}, \bibinfo {author} {\bibfnamefont {X.}~\bibnamefont {Zhang}}, \bibinfo
  {author} {\bibfnamefont {W.}~\bibnamefont {Meng}}, \bibinfo {author}
  {\bibfnamefont {L.}~\bibnamefont {Jin}}, \bibinfo {author} {\bibfnamefont
  {X.}~\bibnamefont {Dai}},\ and\ \bibinfo {author} {\bibfnamefont
  {G.}~\bibnamefont {Liu}},\ }\bibfield  {title} {\bibinfo {title}
  {{Topological nodal lines and nodal points in the antiferromagnetic material
  $\beta$-Fe$_2$PO$_5$}},\ }\href
  {https://doi.org/https://doi.org/10.1039/C9TC04046C} {\bibfield  {journal}
  {\bibinfo  {journal} {J. Mater. Chem. C}\ }\textbf {\bibinfo {volume} {7}},\
  \bibinfo {pages} {12657} (\bibinfo {year} {2019})}\BibitemShut {NoStop}%
\bibitem [{\citenamefont {He}\ \emph {et~al.}(2021)\citenamefont {He},
  \citenamefont {Zhang}, \citenamefont {Liu}, \citenamefont {Dai},
  \citenamefont {Wang},\ and\ \citenamefont {Liu}}]{PhysRevB.104.045143}%
  \BibitemOpen
  \bibfield  {author} {\bibinfo {author} {\bibfnamefont {T.}~\bibnamefont
  {He}}, \bibinfo {author} {\bibfnamefont {X.}~\bibnamefont {Zhang}}, \bibinfo
  {author} {\bibfnamefont {Y.}~\bibnamefont {Liu}}, \bibinfo {author}
  {\bibfnamefont {X.}~\bibnamefont {Dai}}, \bibinfo {author} {\bibfnamefont
  {L.}~\bibnamefont {Wang}},\ and\ \bibinfo {author} {\bibfnamefont
  {G.}~\bibnamefont {Liu}},\ }\bibfield  {title} {\bibinfo {title} {{Potential
  antiferromagnetic Weyl nodal line state in
  ${\mathrm{LiTi}}_{2}{\mathrm{O}}_{4}$ material}},\ }\href
  {https://doi.org/10.1103/PhysRevB.104.045143} {\bibfield  {journal} {\bibinfo
   {journal} {Phys. Rev. B}\ }\textbf {\bibinfo {volume} {104}},\ \bibinfo
  {pages} {045143} (\bibinfo {year} {2021})}\BibitemShut {NoStop}%
\bibitem [{\citenamefont {Zhang}\ \emph
  {et~al.}(2023{\natexlab{a}})\citenamefont {Zhang}, \citenamefont {Wang},
  \citenamefont {Li}, \citenamefont {Zhang}, \citenamefont {Xiao},
  \citenamefont {Liu}, \citenamefont {Luo}, \citenamefont {Lu}, \citenamefont
  {Tian}, \citenamefont {Sun} \emph {et~al.}}]{zhang2023x}%
  \BibitemOpen
  \bibfield  {author} {\bibinfo {author} {\bibfnamefont {S.-S.}\ \bibnamefont
  {Zhang}}, \bibinfo {author} {\bibfnamefont {Z.-A.}\ \bibnamefont {Wang}},
  \bibinfo {author} {\bibfnamefont {B.}~\bibnamefont {Li}}, \bibinfo {author}
  {\bibfnamefont {S.-H.}\ \bibnamefont {Zhang}}, \bibinfo {author}
  {\bibfnamefont {R.-C.}\ \bibnamefont {Xiao}}, \bibinfo {author}
  {\bibfnamefont {L.-X.}\ \bibnamefont {Liu}}, \bibinfo {author} {\bibfnamefont
  {X.}~\bibnamefont {Luo}}, \bibinfo {author} {\bibfnamefont {W.}~\bibnamefont
  {Lu}}, \bibinfo {author} {\bibfnamefont {M.}~\bibnamefont {Tian}}, \bibinfo
  {author} {\bibfnamefont {Y.}~\bibnamefont {Sun}}, \emph {et~al.},\ }\bibfield
   {title} {\bibinfo {title} {{X-Type Antiferromagnets}},\ }\href
  {https://doi.org/10.48550/arXiv.2310.13271} {\bibfield  {journal} {\bibinfo
  {journal} {arXiv:2310.13271}\ } (\bibinfo {year}
  {2023}{\natexlab{a}})}\BibitemShut {NoStop}%
\bibitem [{\citenamefont {Modaressi}\ \emph {et~al.}(1981)\citenamefont
  {Modaressi}, \citenamefont {Courtois}, \citenamefont {Gerardin},
  \citenamefont {Malaman},\ and\ \citenamefont {Gleitzer}}]{MODARESSI1981301}%
  \BibitemOpen
  \bibfield  {author} {\bibinfo {author} {\bibfnamefont {A.}~\bibnamefont
  {Modaressi}}, \bibinfo {author} {\bibfnamefont {A.}~\bibnamefont {Courtois}},
  \bibinfo {author} {\bibfnamefont {R.}~\bibnamefont {Gerardin}}, \bibinfo
  {author} {\bibfnamefont {B.}~\bibnamefont {Malaman}},\ and\ \bibinfo {author}
  {\bibfnamefont {C.}~\bibnamefont {Gleitzer}},\ }\bibfield  {title} {\bibinfo
  {title} {{Fe$_2$PO$_5$, un phosphate de fer de valence mixte. Préparation et
  études structurale, mössbauer et magnétique}},\ }\href
  {https://doi.org/https://doi.org/10.1016/0022-4596(81)90396-0} {\bibfield
  {journal} {\bibinfo  {journal} {J. Solid State Chem.}\ }\textbf {\bibinfo
  {volume} {40}},\ \bibinfo {pages} {301} (\bibinfo {year} {1981})}\BibitemShut
  {NoStop}%
\bibitem [{\citenamefont {Ech-Chahed}\ \emph {et~al.}(1988)\citenamefont
  {Ech-Chahed}, \citenamefont {Jeannot}, \citenamefont {Malaman},\ and\
  \citenamefont {Gleitzer}}]{ECHCHAHED198847}%
  \BibitemOpen
  \bibfield  {author} {\bibinfo {author} {\bibfnamefont {B.}~\bibnamefont
  {Ech-Chahed}}, \bibinfo {author} {\bibfnamefont {F.}~\bibnamefont {Jeannot}},
  \bibinfo {author} {\bibfnamefont {B.}~\bibnamefont {Malaman}},\ and\ \bibinfo
  {author} {\bibfnamefont {C.}~\bibnamefont {Gleitzer}},\ }\bibfield  {title}
  {\bibinfo {title} {{Pre´paration ete´tude d'une varie´te´basse
  tempe´rature de l'oxyphosphate de fer de valence mixte
  $\beta$-Fe$_2$(PO$_4$)O et de NiCr(PO$_4$)O: Un cas d'e´changee´lectronique
  rapide}},\ }\href
  {https://doi.org/https://doi.org/10.1016/0022-4596(88)90330-1} {\bibfield
  {journal} {\bibinfo  {journal} {J. Solid State Chem.}\ }\textbf {\bibinfo
  {volume} {74}},\ \bibinfo {pages} {47} (\bibinfo {year} {1988})}\BibitemShut
  {NoStop}%
\bibitem [{\citenamefont {Ijjaali}\ \emph {et~al.}(1990)\citenamefont
  {Ijjaali}, \citenamefont {Malaman}, \citenamefont {Gleitzer}, \citenamefont
  {Warner}, \citenamefont {Hriljac},\ and\ \citenamefont
  {Cheetham}}]{IJJAALI1990195}%
  \BibitemOpen
  \bibfield  {author} {\bibinfo {author} {\bibfnamefont {M.}~\bibnamefont
  {Ijjaali}}, \bibinfo {author} {\bibfnamefont {B.}~\bibnamefont {Malaman}},
  \bibinfo {author} {\bibfnamefont {C.}~\bibnamefont {Gleitzer}}, \bibinfo
  {author} {\bibfnamefont {J.}~\bibnamefont {Warner}}, \bibinfo {author}
  {\bibfnamefont {J.}~\bibnamefont {Hriljac}},\ and\ \bibinfo {author}
  {\bibfnamefont {A.}~\bibnamefont {Cheetham}},\ }\bibfield  {title} {\bibinfo
  {title} {{Stability, structure refinement, and magnetic properties of
  $\beta$-Fe$_2$(PO$_4$)O}},\ }\href
  {https://doi.org/https://doi.org/10.1016/0022-4596(90)90135-K} {\bibfield
  {journal} {\bibinfo  {journal} {J. Solid State Chem.}\ }\textbf {\bibinfo
  {volume} {86}},\ \bibinfo {pages} {195} (\bibinfo {year} {1990})}\BibitemShut
  {NoStop}%
\bibitem [{\citenamefont {Stokes}\ and\ \citenamefont
  {Hatch}(2005)}]{stokes2005findsym}%
  \BibitemOpen
  \bibfield  {author} {\bibinfo {author} {\bibfnamefont {H.~T.}\ \bibnamefont
  {Stokes}}\ and\ \bibinfo {author} {\bibfnamefont {D.~M.}\ \bibnamefont
  {Hatch}},\ }\bibfield  {title} {\bibinfo {title} {Findsym: program for
  identifying the space-group symmetry of a crystal},\ }\href
  {https://doi.org/https://doi.org/10.1107/S0021889804031528} {\bibfield
  {journal} {\bibinfo  {journal} {J. Appl. Cryst.}\ }\textbf {\bibinfo {volume}
  {38}},\ \bibinfo {pages} {237} (\bibinfo {year} {2005})}\BibitemShut
  {NoStop}%
\bibitem [{SM()}]{SM}%
  \BibitemOpen
  \href@noop {} {}\bibinfo {note} {{See Supplemental Material for the
  calculation methods, the orbital-projected band structure of
  $\beta$-Fe$_2$(PO$_4$)O, the Fermi surfaces of $\beta$-Fe$_2$(PO$_4$)O, the
  band structure of the $\beta$-Fe$_2$(PO$_4$)O with spin-orbit coupling,
  detailed analysis of the Co$_2$(PO$_4$)O compound, band structure and
  electronic conductivity of the lattice model, and the symmetry analysis of
  $Z^3$ nodal net, which includes Refs. \cite{PhysRevB.54.11169, KRESSE199615,
  PhysRevLett.77.3865, PhysRevB.13.5188, PhysRevB.44.943,PhysRevB.57.1505,
  kawamura2019fermisurfer, brinkman1966theory, litvin1974spin, litvin1977spin,
  PhysRevX.12.021016}.}}\BibitemShut {Stop}%
\bibitem [{\citenamefont {Wang}\ \emph {et~al.}(2014)\citenamefont {Wang},
  \citenamefont {Valldor}, \citenamefont {Spielberg},\ and\ \citenamefont
  {Mudring}}]{doi:10.1021/ic4029904}%
  \BibitemOpen
  \bibfield  {author} {\bibinfo {author} {\bibfnamefont {G.}~\bibnamefont
  {Wang}}, \bibinfo {author} {\bibfnamefont {M.}~\bibnamefont {Valldor}},
  \bibinfo {author} {\bibfnamefont {E.~T.}\ \bibnamefont {Spielberg}},\ and\
  \bibinfo {author} {\bibfnamefont {A.-V.}\ \bibnamefont {Mudring}},\
  }\bibfield  {title} {\bibinfo {title} {{Ionothermal Synthesis, Crystal
  Structure, and Magnetic Study of Co$_2$PO$_4$OH Isostructural with
  Caminite}},\ }\href {https://doi.org/10.1021/ic4029904} {\bibfield  {journal}
  {\bibinfo  {journal} {Inorg. Chem.}\ }\textbf {\bibinfo {volume} {53}},\
  \bibinfo {pages} {3072} (\bibinfo {year} {2014})}\BibitemShut {NoStop}%
\bibitem [{\citenamefont {Zhang}\ \emph
  {et~al.}(2022{\natexlab{b}})\citenamefont {Zhang}, \citenamefont {Yu},
  \citenamefont {Liu},\ and\ \citenamefont {Yao}}]{ZHANG2022108153}%
  \BibitemOpen
  \bibfield  {author} {\bibinfo {author} {\bibfnamefont {Z.}~\bibnamefont
  {Zhang}}, \bibinfo {author} {\bibfnamefont {Z.-M.}\ \bibnamefont {Yu}},
  \bibinfo {author} {\bibfnamefont {G.-B.}\ \bibnamefont {Liu}},\ and\ \bibinfo
  {author} {\bibfnamefont {Y.}~\bibnamefont {Yao}},\ }\bibfield  {title}
  {\bibinfo {title} {{MagneticTB: A package for tight-binding model of magnetic
  and non-magnetic materials}},\ }\href
  {https://doi.org/https://doi.org/10.1016/j.cpc.2021.108153} {\bibfield
  {journal} {\bibinfo  {journal} {Comput. Phys. Commun.}\ }\textbf {\bibinfo
  {volume} {270}},\ \bibinfo {pages} {108153} (\bibinfo {year}
  {2022}{\natexlab{b}})}\BibitemShut {NoStop}%
\bibitem [{\citenamefont {Zhang}\ \emph
  {et~al.}(2023{\natexlab{b}})\citenamefont {Zhang}, \citenamefont {Yu},
  \citenamefont {Liu}, \citenamefont {Li}, \citenamefont {Yang},\ and\
  \citenamefont {Yao}}]{zhang2023magnetickp}%
  \BibitemOpen
  \bibfield  {author} {\bibinfo {author} {\bibfnamefont {Z.}~\bibnamefont
  {Zhang}}, \bibinfo {author} {\bibfnamefont {Z.-M.}\ \bibnamefont {Yu}},
  \bibinfo {author} {\bibfnamefont {G.-B.}\ \bibnamefont {Liu}}, \bibinfo
  {author} {\bibfnamefont {Z.}~\bibnamefont {Li}}, \bibinfo {author}
  {\bibfnamefont {S.~A.}\ \bibnamefont {Yang}},\ and\ \bibinfo {author}
  {\bibfnamefont {Y.}~\bibnamefont {Yao}},\ }\bibfield  {title} {\bibinfo
  {title} {{MagneticKP: A package for quickly constructing $ k\cdot p $ models
  of magnetic and non-magnetic crystals}},\ }\href
  {https://doi.org/10.1016/j.cpc.2023.108784} {\bibfield  {journal} {\bibinfo
  {journal} {Comput. Phys. Commun.}\ }\textbf {\bibinfo {volume} {290}},\
  \bibinfo {pages} {108784} (\bibinfo {year} {2023}{\natexlab{b}})}\BibitemShut
  {NoStop}%
\bibitem [{\citenamefont {Wang}(2017{\natexlab{a}})}]{PhysRevB.96.081107}%
  \BibitemOpen
  \bibfield  {author} {\bibinfo {author} {\bibfnamefont {J.}~\bibnamefont
  {Wang}},\ }\bibfield  {title} {\bibinfo {title} {Antiferromagnetic
  topological nodal line semimetals},\ }\href
  {https://doi.org/10.1103/PhysRevB.96.081107} {\bibfield  {journal} {\bibinfo
  {journal} {Phys. Rev. B}\ }\textbf {\bibinfo {volume} {96}},\ \bibinfo
  {pages} {081107} (\bibinfo {year} {2017}{\natexlab{a}})}\BibitemShut
  {NoStop}%
\bibitem [{\citenamefont {Wang}(2017{\natexlab{b}})}]{PhysRevB.95.115138}%
  \BibitemOpen
  \bibfield  {author} {\bibinfo {author} {\bibfnamefont {J.}~\bibnamefont
  {Wang}},\ }\bibfield  {title} {\bibinfo {title} {{Antiferromagnetic Dirac
  semimetals in two dimensions}},\ }\href
  {https://doi.org/10.1103/PhysRevB.95.115138} {\bibfield  {journal} {\bibinfo
  {journal} {Phys. Rev. B}\ }\textbf {\bibinfo {volume} {95}},\ \bibinfo
  {pages} {115138} (\bibinfo {year} {2017}{\natexlab{b}})}\BibitemShut
  {NoStop}%
\bibitem [{\citenamefont {Winkler}\ \emph {et~al.}(2003)\citenamefont
  {Winkler}, \citenamefont {Papadakis}, \citenamefont {De~Poortere},\ and\
  \citenamefont {Shayegan}}]{winkler2003spin}%
  \BibitemOpen
  \bibfield  {author} {\bibinfo {author} {\bibfnamefont {R.}~\bibnamefont
  {Winkler}}, \bibinfo {author} {\bibfnamefont {S.}~\bibnamefont {Papadakis}},
  \bibinfo {author} {\bibfnamefont {E.}~\bibnamefont {De~Poortere}},\ and\
  \bibinfo {author} {\bibfnamefont {M.}~\bibnamefont {Shayegan}},\ }\href@noop
  {} {\emph {\bibinfo {title} {{Spin-orbit coupling in two-dimensional electron
  and hole systems}}}},\ Vol.~\bibinfo {volume} {41}\ (\bibinfo  {publisher}
  {Springer},\ \bibinfo {year} {2003})\BibitemShut {NoStop}%
\bibitem [{\citenamefont {Pizzi}\ \emph {et~al.}(2014)\citenamefont {Pizzi},
  \citenamefont {Volja}, \citenamefont {Kozinsky}, \citenamefont {Fornari},\
  and\ \citenamefont {Marzari}}]{PIZZI2014422}%
  \BibitemOpen
  \bibfield  {author} {\bibinfo {author} {\bibfnamefont {G.}~\bibnamefont
  {Pizzi}}, \bibinfo {author} {\bibfnamefont {D.}~\bibnamefont {Volja}},
  \bibinfo {author} {\bibfnamefont {B.}~\bibnamefont {Kozinsky}}, \bibinfo
  {author} {\bibfnamefont {M.}~\bibnamefont {Fornari}},\ and\ \bibinfo {author}
  {\bibfnamefont {N.}~\bibnamefont {Marzari}},\ }\bibfield  {title} {\bibinfo
  {title} {{BoltzWann: A code for the evaluation of thermoelectric and
  electronic transport properties with a maximally-localized Wannier functions
  basis}},\ }\href {https://doi.org/https://doi.org/10.1016/j.cpc.2013.09.015}
  {\bibfield  {journal} {\bibinfo  {journal} {Comput. Phys. Commun.}\ }\textbf
  {\bibinfo {volume} {185}},\ \bibinfo {pages} {422} (\bibinfo {year}
  {2014})}\BibitemShut {NoStop}%
\bibitem [{\citenamefont {Ziman}(1972)}]{ziman1972principles}%
  \BibitemOpen
  \bibfield  {author} {\bibinfo {author} {\bibfnamefont {J.~M.}\ \bibnamefont
  {Ziman}},\ }\href@noop {} {\emph {\bibinfo {title} {Principles of the Theory
  of Solids}}}\ (\bibinfo  {publisher} {Cambridge university press},\ \bibinfo
  {year} {1972})\BibitemShut {NoStop}%
\bibitem [{\citenamefont {Scheidemantel}\ \emph {et~al.}(2003)\citenamefont
  {Scheidemantel}, \citenamefont {Ambrosch-Draxl}, \citenamefont {Thonhauser},
  \citenamefont {Badding},\ and\ \citenamefont {Sofo}}]{PhysRevB.68.125210}%
  \BibitemOpen
  \bibfield  {author} {\bibinfo {author} {\bibfnamefont {T.~J.}\ \bibnamefont
  {Scheidemantel}}, \bibinfo {author} {\bibfnamefont {C.}~\bibnamefont
  {Ambrosch-Draxl}}, \bibinfo {author} {\bibfnamefont {T.}~\bibnamefont
  {Thonhauser}}, \bibinfo {author} {\bibfnamefont {J.~V.}\ \bibnamefont
  {Badding}},\ and\ \bibinfo {author} {\bibfnamefont {J.~O.}\ \bibnamefont
  {Sofo}},\ }\bibfield  {title} {\bibinfo {title} {Transport coefficients from
  first-principles calculations},\ }\href
  {https://doi.org/10.1103/PhysRevB.68.125210} {\bibfield  {journal} {\bibinfo
  {journal} {Phys. Rev. B}\ }\textbf {\bibinfo {volume} {68}},\ \bibinfo
  {pages} {125210} (\bibinfo {year} {2003})}\BibitemShut {NoStop}%
\bibitem [{\citenamefont {Madsen}(2006)}]{madsen2006automated}%
  \BibitemOpen
  \bibfield  {author} {\bibinfo {author} {\bibfnamefont {G.~K.~H.}\
  \bibnamefont {Madsen}},\ }\bibfield  {title} {\bibinfo {title} {Automated
  search for new thermoelectric materials: The case of liznsb},\ }\href
  {https://doi.org/10.1021/ja062526a} {\bibfield  {journal} {\bibinfo
  {journal} {J. Am. Chem. Soc.}\ }\textbf {\bibinfo {volume} {128}},\ \bibinfo
  {pages} {12140} (\bibinfo {year} {2006})}\BibitemShut {NoStop}%
\bibitem [{\citenamefont {Nag}(2012)}]{nag2012electron}%
  \BibitemOpen
  \bibfield  {author} {\bibinfo {author} {\bibfnamefont {B.~R.}\ \bibnamefont
  {Nag}},\ }\href@noop {} {\emph {\bibinfo {title} {Electron transport in
  compound semiconductors}}},\ Vol.~\bibinfo {volume} {11}\ (\bibinfo
  {publisher} {Springer Science \& Business Media},\ \bibinfo {year}
  {2012})\BibitemShut {NoStop}%
\bibitem [{\citenamefont {Mostofi}\ \emph {et~al.}(2008)\citenamefont
  {Mostofi}, \citenamefont {Yates}, \citenamefont {Lee}, \citenamefont {Souza},
  \citenamefont {Vanderbilt},\ and\ \citenamefont
  {Marzari}}]{mostofi2008wannier90}%
  \BibitemOpen
  \bibfield  {author} {\bibinfo {author} {\bibfnamefont {A.~A.}\ \bibnamefont
  {Mostofi}}, \bibinfo {author} {\bibfnamefont {J.~R.}\ \bibnamefont {Yates}},
  \bibinfo {author} {\bibfnamefont {Y.-S.}\ \bibnamefont {Lee}}, \bibinfo
  {author} {\bibfnamefont {I.}~\bibnamefont {Souza}}, \bibinfo {author}
  {\bibfnamefont {D.}~\bibnamefont {Vanderbilt}},\ and\ \bibinfo {author}
  {\bibfnamefont {N.}~\bibnamefont {Marzari}},\ }\bibfield  {title} {\bibinfo
  {title} {Wannier90: {{A}} tool for obtaining maximally-localised {{Wannier}}
  functions},\ }\href {https://doi.org/10.1016/j.cpc.2007.11.016} {\bibfield
  {journal} {\bibinfo  {journal} {Comput. Phys. Commun.}\ }\textbf {\bibinfo
  {volume} {178}},\ \bibinfo {pages} {685} (\bibinfo {year}
  {2008})}\BibitemShut {NoStop}%
\bibitem [{\citenamefont {Mostofi}\ \emph {et~al.}(2014)\citenamefont
  {Mostofi}, \citenamefont {Yates}, \citenamefont {Pizzi}, \citenamefont {Lee},
  \citenamefont {Souza}, \citenamefont {Vanderbilt},\ and\ \citenamefont
  {Marzari}}]{mostofi2014updated}%
  \BibitemOpen
  \bibfield  {author} {\bibinfo {author} {\bibfnamefont {A.~A.}\ \bibnamefont
  {Mostofi}}, \bibinfo {author} {\bibfnamefont {J.~R.}\ \bibnamefont {Yates}},
  \bibinfo {author} {\bibfnamefont {G.}~\bibnamefont {Pizzi}}, \bibinfo
  {author} {\bibfnamefont {Y.-S.}\ \bibnamefont {Lee}}, \bibinfo {author}
  {\bibfnamefont {I.}~\bibnamefont {Souza}}, \bibinfo {author} {\bibfnamefont
  {D.}~\bibnamefont {Vanderbilt}},\ and\ \bibinfo {author} {\bibfnamefont
  {N.}~\bibnamefont {Marzari}},\ }\bibfield  {title} {\bibinfo {title} {An
  updated version of wannier90: {{A}} tool for obtaining maximally-localised
  {{Wannier}} functions},\ }\href {https://doi.org/10.1016/j.cpc.2014.05.003}
  {\bibfield  {journal} {\bibinfo  {journal} {Comput. Phys. Commun.}\ }\textbf
  {\bibinfo {volume} {185}},\ \bibinfo {pages} {2309} (\bibinfo {year}
  {2014})}\BibitemShut {NoStop}%
\bibitem [{\citenamefont {Feng}\ \emph {et~al.}(2022)\citenamefont {Feng},
  \citenamefont {Zhou}, \citenamefont {Smejkal}, \citenamefont {Wu},
  \citenamefont {Zhu}, \citenamefont {Guo}, \citenamefont {Gonzalez-Hernandez},
  \citenamefont {Wang}, \citenamefont {Yan}, \citenamefont {Qin}, \citenamefont
  {Zhang}, \citenamefont {Wu}, \citenamefont {Chen}, \citenamefont {Meng},
  \citenamefont {Liu}, \citenamefont {Xia}, \citenamefont {Sinova},
  \citenamefont {Jungwirth},\ and\ \citenamefont {Liu}}]{feng2022anomalous}%
  \BibitemOpen
  \bibfield  {author} {\bibinfo {author} {\bibfnamefont {Z.}~\bibnamefont
  {Feng}}, \bibinfo {author} {\bibfnamefont {X.}~\bibnamefont {Zhou}}, \bibinfo
  {author} {\bibfnamefont {L.}~\bibnamefont {Smejkal}}, \bibinfo {author}
  {\bibfnamefont {L.}~\bibnamefont {Wu}}, \bibinfo {author} {\bibfnamefont
  {Z.}~\bibnamefont {Zhu}}, \bibinfo {author} {\bibfnamefont {H.}~\bibnamefont
  {Guo}}, \bibinfo {author} {\bibfnamefont {R.}~\bibnamefont
  {Gonzalez-Hernandez}}, \bibinfo {author} {\bibfnamefont {X.}~\bibnamefont
  {Wang}}, \bibinfo {author} {\bibfnamefont {H.}~\bibnamefont {Yan}}, \bibinfo
  {author} {\bibfnamefont {P.}~\bibnamefont {Qin}}, \bibinfo {author}
  {\bibfnamefont {X.}~\bibnamefont {Zhang}}, \bibinfo {author} {\bibfnamefont
  {H.}~\bibnamefont {Wu}}, \bibinfo {author} {\bibfnamefont {H.}~\bibnamefont
  {Chen}}, \bibinfo {author} {\bibfnamefont {Z.}~\bibnamefont {Meng}}, \bibinfo
  {author} {\bibfnamefont {L.}~\bibnamefont {Liu}}, \bibinfo {author}
  {\bibfnamefont {Z.}~\bibnamefont {Xia}}, \bibinfo {author} {\bibfnamefont
  {J.}~\bibnamefont {Sinova}}, \bibinfo {author} {\bibfnamefont
  {T.}~\bibnamefont {Jungwirth}},\ and\ \bibinfo {author} {\bibfnamefont
  {Z.}~\bibnamefont {Liu}},\ }\bibfield  {title} {\bibinfo {title} {An
  anomalous hall effect in altermagnetic ruthenium dioxide},\ }\href
  {https://doi.org/10.1038/s41928-022-00866-z} {\bibfield  {journal} {\bibinfo
  {journal} {Nat Electron}\ }\textbf {\bibinfo {volume} {5}},\ \bibinfo {pages}
  {735} (\bibinfo {year} {2022})}\BibitemShut {NoStop}%
\bibitem [{\citenamefont {Kresse}\ and\ \citenamefont
  {Furthm\"uller}(1996)}]{PhysRevB.54.11169}%
  \BibitemOpen
  \bibfield  {author} {\bibinfo {author} {\bibfnamefont {G.}~\bibnamefont
  {Kresse}}\ and\ \bibinfo {author} {\bibfnamefont {J.}~\bibnamefont
  {Furthm\"uller}},\ }\bibfield  {title} {\bibinfo {title} {{Efficient
  iterative schemes for ab initio total-energy calculations using a plane-wave
  basis set}},\ }\href {https://doi.org/10.1103/PhysRevB.54.11169} {\bibfield
  {journal} {\bibinfo  {journal} {Phys. Rev. B}\ }\textbf {\bibinfo {volume}
  {54}},\ \bibinfo {pages} {11169} (\bibinfo {year} {1996})}\BibitemShut
  {NoStop}%
\bibitem [{\citenamefont {Kresse}\ and\ \citenamefont
  {Furthmüller}(1996)}]{KRESSE199615}%
  \BibitemOpen
  \bibfield  {author} {\bibinfo {author} {\bibfnamefont {G.}~\bibnamefont
  {Kresse}}\ and\ \bibinfo {author} {\bibfnamefont {J.}~\bibnamefont
  {Furthmüller}},\ }\bibfield  {title} {\bibinfo {title} {{Efficiency of
  ab-initio total energy calculations for metals and semiconductors using a
  plane-wave basis set}},\ }\href
  {https://doi.org/https://doi.org/10.1016/0927-0256(96)00008-0} {\bibfield
  {journal} {\bibinfo  {journal} {Comp. Mater. Sci.}\ }\textbf {\bibinfo
  {volume} {6}},\ \bibinfo {pages} {15} (\bibinfo {year} {1996})}\BibitemShut
  {NoStop}%
\bibitem [{\citenamefont {Perdew}\ \emph {et~al.}(1996)\citenamefont {Perdew},
  \citenamefont {Burke},\ and\ \citenamefont
  {Ernzerhof}}]{PhysRevLett.77.3865}%
  \BibitemOpen
  \bibfield  {author} {\bibinfo {author} {\bibfnamefont {J.~P.}\ \bibnamefont
  {Perdew}}, \bibinfo {author} {\bibfnamefont {K.}~\bibnamefont {Burke}},\ and\
  \bibinfo {author} {\bibfnamefont {M.}~\bibnamefont {Ernzerhof}},\ }\bibfield
  {title} {\bibinfo {title} {{Generalized Gradient Approximation Made
  Simple}},\ }\href {https://doi.org/10.1103/PhysRevLett.77.3865} {\bibfield
  {journal} {\bibinfo  {journal} {Phys. Rev. Lett.}\ }\textbf {\bibinfo
  {volume} {77}},\ \bibinfo {pages} {3865} (\bibinfo {year}
  {1996})}\BibitemShut {NoStop}%
\bibitem [{\citenamefont {Monkhorst}\ and\ \citenamefont
  {Pack}(1976)}]{PhysRevB.13.5188}%
  \BibitemOpen
  \bibfield  {author} {\bibinfo {author} {\bibfnamefont {H.~J.}\ \bibnamefont
  {Monkhorst}}\ and\ \bibinfo {author} {\bibfnamefont {J.~D.}\ \bibnamefont
  {Pack}},\ }\bibfield  {title} {\bibinfo {title} {{Special points for
  Brillouin-zone integrations}},\ }\href
  {https://doi.org/10.1103/PhysRevB.13.5188} {\bibfield  {journal} {\bibinfo
  {journal} {Phys. Rev. B}\ }\textbf {\bibinfo {volume} {13}},\ \bibinfo
  {pages} {5188} (\bibinfo {year} {1976})}\BibitemShut {NoStop}%
\bibitem [{\citenamefont {Anisimov}\ \emph {et~al.}(1991)\citenamefont
  {Anisimov}, \citenamefont {Zaanen},\ and\ \citenamefont
  {Andersen}}]{PhysRevB.44.943}%
  \BibitemOpen
  \bibfield  {author} {\bibinfo {author} {\bibfnamefont {V.~I.}\ \bibnamefont
  {Anisimov}}, \bibinfo {author} {\bibfnamefont {J.}~\bibnamefont {Zaanen}},\
  and\ \bibinfo {author} {\bibfnamefont {O.~K.}\ \bibnamefont {Andersen}},\
  }\bibfield  {title} {\bibinfo {title} {{Band theory and Mott insulators:
  Hubbard U instead of Stoner I}},\ }\href
  {https://doi.org/10.1103/PhysRevB.44.943} {\bibfield  {journal} {\bibinfo
  {journal} {Phys. Rev. B}\ }\textbf {\bibinfo {volume} {44}},\ \bibinfo
  {pages} {943} (\bibinfo {year} {1991})}\BibitemShut {NoStop}%
\bibitem [{\citenamefont {Dudarev}\ \emph {et~al.}(1998)\citenamefont
  {Dudarev}, \citenamefont {Botton}, \citenamefont {Savrasov}, \citenamefont
  {Humphreys},\ and\ \citenamefont {Sutton}}]{PhysRevB.57.1505}%
  \BibitemOpen
  \bibfield  {author} {\bibinfo {author} {\bibfnamefont {S.~L.}\ \bibnamefont
  {Dudarev}}, \bibinfo {author} {\bibfnamefont {G.~A.}\ \bibnamefont {Botton}},
  \bibinfo {author} {\bibfnamefont {S.~Y.}\ \bibnamefont {Savrasov}}, \bibinfo
  {author} {\bibfnamefont {C.~J.}\ \bibnamefont {Humphreys}},\ and\ \bibinfo
  {author} {\bibfnamefont {A.~P.}\ \bibnamefont {Sutton}},\ }\bibfield  {title}
  {\bibinfo {title} {{Electron-energy-loss spectra and the structural stability
  of nickel oxide: An LSDA+U study}},\ }\href
  {https://doi.org/10.1103/PhysRevB.57.1505} {\bibfield  {journal} {\bibinfo
  {journal} {Phys. Rev. B}\ }\textbf {\bibinfo {volume} {57}},\ \bibinfo
  {pages} {1505} (\bibinfo {year} {1998})}\BibitemShut {NoStop}%
\bibitem [{\citenamefont {Kawamura}(2019)}]{kawamura2019fermisurfer}%
  \BibitemOpen
  \bibfield  {author} {\bibinfo {author} {\bibfnamefont {M.}~\bibnamefont
  {Kawamura}},\ }\bibfield  {title} {\bibinfo {title} {{FermiSurfer:
  Fermi-surface viewer providing multiple representation schemes}},\ }\href
  {https://doi.org/https://doi.org/10.1016/j.cpc.2019.01.017} {\bibfield
  {journal} {\bibinfo  {journal} {Comput. Phys. Commun.}\ }\textbf {\bibinfo
  {volume} {239}},\ \bibinfo {pages} {197} (\bibinfo {year}
  {2019})}\BibitemShut {NoStop}%
\bibitem [{\citenamefont {Brinkman}\ and\ \citenamefont
  {Elliott}(1966)}]{brinkman1966theory}%
  \BibitemOpen
  \bibfield  {author} {\bibinfo {author} {\bibfnamefont {W.}~\bibnamefont
  {Brinkman}}\ and\ \bibinfo {author} {\bibfnamefont {R.~J.}\ \bibnamefont
  {Elliott}},\ }\bibfield  {title} {\bibinfo {title} {{Theory of spin-space
  groups}},\ }\href {https://doi.org/https://doi.org/10.1098/rspa.1966.0211}
  {\bibfield  {journal} {\bibinfo  {journal} {Proceedings of the Royal Society
  of London. Series A. Mathematical and Physical Sciences}\ }\textbf {\bibinfo
  {volume} {294}},\ \bibinfo {pages} {343} (\bibinfo {year}
  {1966})}\BibitemShut {NoStop}%
\bibitem [{\citenamefont {Litvin}\ and\ \citenamefont
  {Opechowski}(1974)}]{litvin1974spin}%
  \BibitemOpen
  \bibfield  {author} {\bibinfo {author} {\bibfnamefont {D.~B.}\ \bibnamefont
  {Litvin}}\ and\ \bibinfo {author} {\bibfnamefont {W.}~\bibnamefont
  {Opechowski}},\ }\bibfield  {title} {\bibinfo {title} {{Spin groups}},\
  }\href {https://doi.org/https://doi.org/10.1016/0031-8914(74)90157-8}
  {\bibfield  {journal} {\bibinfo  {journal} {Physica}\ }\textbf {\bibinfo
  {volume} {76}},\ \bibinfo {pages} {538} (\bibinfo {year} {1974})}\BibitemShut
  {NoStop}%
\bibitem [{\citenamefont {Litvin}(1977)}]{litvin1977spin}%
  \BibitemOpen
  \bibfield  {author} {\bibinfo {author} {\bibfnamefont {D.~B.}\ \bibnamefont
  {Litvin}},\ }\bibfield  {title} {\bibinfo {title} {{Spin point groups}},\
  }\href {https://doi.org/https://doi.org/10.1107/S0567739477000709} {\bibfield
   {journal} {\bibinfo  {journal} {Acta Crystallographica Section A: Crystal
  Physics, Diffraction, Theoretical and General Crystallography}\ }\textbf
  {\bibinfo {volume} {33}},\ \bibinfo {pages} {279} (\bibinfo {year}
  {1977})}\BibitemShut {NoStop}%
\bibitem [{\citenamefont {Liu}\ \emph {et~al.}(2022{\natexlab{b}})\citenamefont
  {Liu}, \citenamefont {Li}, \citenamefont {Han}, \citenamefont {Wan},\ and\
  \citenamefont {Liu}}]{PhysRevX.12.021016}%
  \BibitemOpen
  \bibfield  {author} {\bibinfo {author} {\bibfnamefont {P.}~\bibnamefont
  {Liu}}, \bibinfo {author} {\bibfnamefont {J.}~\bibnamefont {Li}}, \bibinfo
  {author} {\bibfnamefont {J.}~\bibnamefont {Han}}, \bibinfo {author}
  {\bibfnamefont {X.}~\bibnamefont {Wan}},\ and\ \bibinfo {author}
  {\bibfnamefont {Q.}~\bibnamefont {Liu}},\ }\bibfield  {title} {\bibinfo
  {title} {{Spin-Group Symmetry in Magnetic Materials with Negligible
  Spin-Orbit Coupling}},\ }\href {https://doi.org/10.1103/PhysRevX.12.021016}
  {\bibfield  {journal} {\bibinfo  {journal} {Phys. Rev. X}\ }\textbf {\bibinfo
  {volume} {12}},\ \bibinfo {pages} {021016} (\bibinfo {year}
  {2022}{\natexlab{b}})}\BibitemShut {NoStop}%
\end{thebibliography}%


\begin{thebibliography}{11}%
\makeatletter
\providecommand \@ifxundefined [1]{%
 \@ifx{#1\undefined}
}%
\providecommand \@ifnum [1]{%
 \ifnum #1\expandafter \@firstoftwo
 \else \expandafter \@secondoftwo
 \fi
}%
\providecommand \@ifx [1]{%
 \ifx #1\expandafter \@firstoftwo
 \else \expandafter \@secondoftwo
 \fi
}%
\providecommand \natexlab [1]{#1}%
\providecommand \enquote  [1]{``#1''}%
\providecommand \bibnamefont  [1]{#1}%
\providecommand \bibfnamefont [1]{#1}%
\providecommand \citenamefont [1]{#1}%
\providecommand \href@noop [0]{\@secondoftwo}%
\providecommand \href [0]{\begingroup \@sanitize@url \@href}%
\providecommand \@href[1]{\@@startlink{#1}\@@href}%
\providecommand \@@href[1]{\endgroup#1\@@endlink}%
\providecommand \@sanitize@url [0]{\catcode `\\12\catcode `\$12\catcode
  `\&12\catcode `\#12\catcode `\^12\catcode `\_12\catcode `\%12\relax}%
\providecommand \@@startlink[1]{}%
\providecommand \@@endlink[0]{}%
\providecommand \url  [0]{\begingroup\@sanitize@url \@url }%
\providecommand \@url [1]{\endgroup\@href {#1}{\urlprefix }}%
\providecommand \urlprefix  [0]{URL }%
\providecommand \Eprint [0]{\href }%
\providecommand \doibase [0]{https://doi.org/}%
\providecommand \selectlanguage [0]{\@gobble}%
\providecommand \bibinfo  [0]{\@secondoftwo}%
\providecommand \bibfield  [0]{\@secondoftwo}%
\providecommand \translation [1]{[#1]}%
\providecommand \BibitemOpen [0]{}%
\providecommand \bibitemStop [0]{}%
\providecommand \bibitemNoStop [0]{.\EOS\space}%
\providecommand \EOS [0]{\spacefactor3000\relax}%
\providecommand \BibitemShut  [1]{\csname bibitem#1\endcsname}%
\let\auto@bib@innerbib\@empty
\bibitem [{\citenamefont {Kresse}\ and\ \citenamefont
  {Furthm\"uller}(1996)}]{PhysRevB.54.11169}%
  \BibitemOpen
  \bibfield  {author} {\bibinfo {author} {\bibfnamefont {G.}~\bibnamefont
  {Kresse}}\ and\ \bibinfo {author} {\bibfnamefont {J.}~\bibnamefont
  {Furthm\"uller}},\ }\bibfield  {title} {\bibinfo {title} {{Efficient
  iterative schemes for ab initio total-energy calculations using a plane-wave
  basis set}},\ }\href {https://doi.org/10.1103/PhysRevB.54.11169} {\bibfield
  {journal} {\bibinfo  {journal} {Phys. Rev. B}\ }\textbf {\bibinfo {volume}
  {54}},\ \bibinfo {pages} {11169} (\bibinfo {year} {1996})}\BibitemShut
  {NoStop}%
\bibitem [{\citenamefont {Kresse}\ and\ \citenamefont
  {Furthmüller}(1996)}]{KRESSE199615}%
  \BibitemOpen
  \bibfield  {author} {\bibinfo {author} {\bibfnamefont {G.}~\bibnamefont
  {Kresse}}\ and\ \bibinfo {author} {\bibfnamefont {J.}~\bibnamefont
  {Furthmüller}},\ }\bibfield  {title} {\bibinfo {title} {{Efficiency of
  ab-initio total energy calculations for metals and semiconductors using a
  plane-wave basis set}},\ }\href
  {https://doi.org/https://doi.org/10.1016/0927-0256(96)00008-0} {\bibfield
  {journal} {\bibinfo  {journal} {Comp. Mater. Sci.}\ }\textbf {\bibinfo
  {volume} {6}},\ \bibinfo {pages} {15} (\bibinfo {year} {1996})}\BibitemShut
  {NoStop}%
\bibitem [{\citenamefont {Perdew}\ \emph {et~al.}(1996)\citenamefont {Perdew},
  \citenamefont {Burke},\ and\ \citenamefont
  {Ernzerhof}}]{PhysRevLett.77.3865}%
  \BibitemOpen
  \bibfield  {author} {\bibinfo {author} {\bibfnamefont {J.~P.}\ \bibnamefont
  {Perdew}}, \bibinfo {author} {\bibfnamefont {K.}~\bibnamefont {Burke}},\ and\
  \bibinfo {author} {\bibfnamefont {M.}~\bibnamefont {Ernzerhof}},\ }\bibfield
  {title} {\bibinfo {title} {{Generalized Gradient Approximation Made
  Simple}},\ }\href {https://doi.org/10.1103/PhysRevLett.77.3865} {\bibfield
  {journal} {\bibinfo  {journal} {Phys. Rev. Lett.}\ }\textbf {\bibinfo
  {volume} {77}},\ \bibinfo {pages} {3865} (\bibinfo {year}
  {1996})}\BibitemShut {NoStop}%
\bibitem [{\citenamefont {Monkhorst}\ and\ \citenamefont
  {Pack}(1976)}]{PhysRevB.13.5188}%
  \BibitemOpen
  \bibfield  {author} {\bibinfo {author} {\bibfnamefont {H.~J.}\ \bibnamefont
  {Monkhorst}}\ and\ \bibinfo {author} {\bibfnamefont {J.~D.}\ \bibnamefont
  {Pack}},\ }\bibfield  {title} {\bibinfo {title} {Special points for
  brillouin-zone integrations},\ }\href
  {https://doi.org/10.1103/PhysRevB.13.5188} {\bibfield  {journal} {\bibinfo
  {journal} {Phys. Rev. B}\ }\textbf {\bibinfo {volume} {13}},\ \bibinfo
  {pages} {5188} (\bibinfo {year} {1976})}\BibitemShut {NoStop}%
\bibitem [{\citenamefont {Anisimov}\ \emph {et~al.}(1991)\citenamefont
  {Anisimov}, \citenamefont {Zaanen},\ and\ \citenamefont
  {Andersen}}]{PhysRevB.44.943}%
  \BibitemOpen
  \bibfield  {author} {\bibinfo {author} {\bibfnamefont {V.~I.}\ \bibnamefont
  {Anisimov}}, \bibinfo {author} {\bibfnamefont {J.}~\bibnamefont {Zaanen}},\
  and\ \bibinfo {author} {\bibfnamefont {O.~K.}\ \bibnamefont {Andersen}},\
  }\bibfield  {title} {\bibinfo {title} {{Band theory and Mott insulators:
  Hubbard U instead of Stoner I}},\ }\href
  {https://doi.org/10.1103/PhysRevB.44.943} {\bibfield  {journal} {\bibinfo
  {journal} {Phys. Rev. B}\ }\textbf {\bibinfo {volume} {44}},\ \bibinfo
  {pages} {943} (\bibinfo {year} {1991})}\BibitemShut {NoStop}%
\bibitem [{\citenamefont {Dudarev}\ \emph {et~al.}(1998)\citenamefont
  {Dudarev}, \citenamefont {Botton}, \citenamefont {Savrasov}, \citenamefont
  {Humphreys},\ and\ \citenamefont {Sutton}}]{PhysRevB.57.1505}%
  \BibitemOpen
  \bibfield  {author} {\bibinfo {author} {\bibfnamefont {S.~L.}\ \bibnamefont
  {Dudarev}}, \bibinfo {author} {\bibfnamefont {G.~A.}\ \bibnamefont {Botton}},
  \bibinfo {author} {\bibfnamefont {S.~Y.}\ \bibnamefont {Savrasov}}, \bibinfo
  {author} {\bibfnamefont {C.~J.}\ \bibnamefont {Humphreys}},\ and\ \bibinfo
  {author} {\bibfnamefont {A.~P.}\ \bibnamefont {Sutton}},\ }\bibfield  {title}
  {\bibinfo {title} {{Electron-energy-loss spectra and the structural stability
  of nickel oxide: An LSDA+U study}},\ }\href
  {https://doi.org/10.1103/PhysRevB.57.1505} {\bibfield  {journal} {\bibinfo
  {journal} {Phys. Rev. B}\ }\textbf {\bibinfo {volume} {57}},\ \bibinfo
  {pages} {1505} (\bibinfo {year} {1998})}\BibitemShut {NoStop}%
\bibitem [{\citenamefont {Kawamura}(2019)}]{kawamura2019fermisurfer}%
  \BibitemOpen
  \bibfield  {author} {\bibinfo {author} {\bibfnamefont {M.}~\bibnamefont
  {Kawamura}},\ }\bibfield  {title} {\bibinfo {title} {{FermiSurfer:
  Fermi-surface viewer providing multiple representation schemes}},\ }\href
  {https://doi.org/https://doi.org/10.1016/j.cpc.2019.01.017} {\bibfield
  {journal} {\bibinfo  {journal} {Comput. Phys. Commun.}\ }\textbf {\bibinfo
  {volume} {239}},\ \bibinfo {pages} {197} (\bibinfo {year}
  {2019})}\BibitemShut {NoStop}%
\bibitem [{\citenamefont {Brinkman}\ and\ \citenamefont
  {Elliott}(1966)}]{brinkman1966theory}%
  \BibitemOpen
  \bibfield  {author} {\bibinfo {author} {\bibfnamefont {W.}~\bibnamefont
  {Brinkman}}\ and\ \bibinfo {author} {\bibfnamefont {R.~J.}\ \bibnamefont
  {Elliott}},\ }\bibfield  {title} {\bibinfo {title} {{Theory of spin-space
  groups}},\ }\href {https://doi.org/https://doi.org/10.1098/rspa.1966.0211}
  {\bibfield  {journal} {\bibinfo  {journal} {Proceedings of the Royal Society
  of London. Series A. Mathematical and Physical Sciences}\ }\textbf {\bibinfo
  {volume} {294}},\ \bibinfo {pages} {343} (\bibinfo {year}
  {1966})}\BibitemShut {NoStop}%
\bibitem [{\citenamefont {Litvin}\ and\ \citenamefont
  {Opechowski}(1974)}]{litvin1974spin}%
  \BibitemOpen
  \bibfield  {author} {\bibinfo {author} {\bibfnamefont {D.~B.}\ \bibnamefont
  {Litvin}}\ and\ \bibinfo {author} {\bibfnamefont {W.}~\bibnamefont
  {Opechowski}},\ }\bibfield  {title} {\bibinfo {title} {{Spin groups}},\
  }\href {https://doi.org/https://doi.org/10.1016/0031-8914(74)90157-8}
  {\bibfield  {journal} {\bibinfo  {journal} {Physica}\ }\textbf {\bibinfo
  {volume} {76}},\ \bibinfo {pages} {538} (\bibinfo {year} {1974})}\BibitemShut
  {NoStop}%
\bibitem [{\citenamefont {Litvin}(1977)}]{litvin1977spin}%
  \BibitemOpen
  \bibfield  {author} {\bibinfo {author} {\bibfnamefont {D.~B.}\ \bibnamefont
  {Litvin}},\ }\bibfield  {title} {\bibinfo {title} {{Spin point groups}},\
  }\href {https://doi.org/https://doi.org/10.1107/S0567739477000709} {\bibfield
   {journal} {\bibinfo  {journal} {Acta Crystallographica Section A: Crystal
  Physics, Diffraction, Theoretical and General Crystallography}\ }\textbf
  {\bibinfo {volume} {33}},\ \bibinfo {pages} {279} (\bibinfo {year}
  {1977})}\BibitemShut {NoStop}%
\bibitem [{\citenamefont {Liu}\ \emph {et~al.}(2022)\citenamefont {Liu},
  \citenamefont {Li}, \citenamefont {Han}, \citenamefont {Wan},\ and\
  \citenamefont {Liu}}]{PhysRevX.12.021016}%
  \BibitemOpen
  \bibfield  {author} {\bibinfo {author} {\bibfnamefont {P.}~\bibnamefont
  {Liu}}, \bibinfo {author} {\bibfnamefont {J.}~\bibnamefont {Li}}, \bibinfo
  {author} {\bibfnamefont {J.}~\bibnamefont {Han}}, \bibinfo {author}
  {\bibfnamefont {X.}~\bibnamefont {Wan}},\ and\ \bibinfo {author}
  {\bibfnamefont {Q.}~\bibnamefont {Liu}},\ }\bibfield  {title} {\bibinfo
  {title} {{Spin-Group Symmetry in Magnetic Materials with Negligible
  Spin-Orbit Coupling}},\ }\href {https://doi.org/10.1103/PhysRevX.12.021016}
  {\bibfield  {journal} {\bibinfo  {journal} {Phys. Rev. X}\ }\textbf {\bibinfo
  {volume} {12}},\ \bibinfo {pages} {021016} (\bibinfo {year}
  {2022})}\BibitemShut {NoStop}%
\end{thebibliography}%

\end{document}